\documentclass[aps,prb,reprint,amsmath,amssymb,floatfix,superscriptaddress]{revtex4-2}
\usepackage[usenames]{color}
\usepackage{graphicx}
\usepackage{dcolumn}
\usepackage{bm}
\usepackage{hyperref,epsfig,wrapfig}
\usepackage{amsmath}

\begin{document}

\title{Metamagnetism and tricriticalilty in heavy-fermion ferromagnet URhGe}

\author{A. S. Gubina}
\affiliation{Universit\'e Grenoble Alpes, CEA, IRIG, PHELIQS, 38000 Grenoble, France}
\author{M. E. Zhitomirsky}
\affiliation{Universit\'e Grenoble Alpes, CEA, IRIG, PHELIQS, 38000 Grenoble, France}
\affiliation{Institut Laue Langevin, 38042 Grenoble Cedex 9, France}
\date{\today}

\begin{abstract}
URhGe is a ferromagnetic superconductor with a distinctive magnetic behavior. 
In a field $H\parallel b$ applied perpendicular to the easy  axis,
URhGe exhibits an abrupt orientational transition of the magnetization with
a reentrant superconducting phase emerging close to the transition field $H_m$.  
We develop a theoretical description of the magnetic properties of URhGe by 
considering a spin model with competing magnetic anisotropies. The model is analyzed both 
analytically at zero temperature and with Monte Carlo simulations at finite temperatures. 
The constructed $H_b$--$T$ phase diagram features a tricritical point on the line $H_m(T)$ and
 is in a good quantitative agreement with the experimental diagram of URhGe.
We demonstrate that the asymptotic tricritical behavior of the order parameter and the correlation
length is described by  the mean-field critical exponents.  The derived microscopic parameters
suggest that URhGe is an $XY$-like ferromagnet with an additional weak in-plane anisotropy in the $bc$ plane.
\end{abstract}
\maketitle

%%%%%%%%%%%%%%%%%%%%%%%%%%%%%%%%%%%%%%%%%%%%%%%%%%%
\section{Introduction}
%%%%%%%%%%%%%%%%%%%%%%%%%%%%%%%%%%%%%%%%%%%%%%%%%%%

It is generally accepted that ferromagnetism is detrimental to superconductivity. 
The discovery of coexisting superconducting and ferromagnetic phases 
in the heavy fermion materials UGe$_2$ \cite{Saxena00}, URhGe \cite{Aoki01}, and UCoGe  \cite{Huy07} 
challenges this common perception.  The exact mechanism responsible for the simultaneous presence of two antagonistic 
states is still a matter of debate, see \cite{Sandeman03,Nevidomsky05,Karchev03,Mineev11,Hattori13,Bulaevskii19,Mineev20} 
for  theoretical discussions  and  \cite{Huxley15,Mineev17,Aoki19} for general overviews.
To make progress, a deeper understanding of the magnetic properties of these uranium compounds
is required.
In our work, we focus on URhGe, in which superconductivity and  ferromagnetism are present at ambient pressure.
Interestingly, URhGe has a second superconducting pocket in a strong magnetic field $H\parallel b$  \cite{Levy05}.  The reentrant superconducting phase resides in the vicinity of the metamagnetic  
transition at $H_m = 11.7$~T, which corresponds to a discontinuous rotation of the ferromagnetic moment   
from a tilted orientation to the field direction. 

Theoretical description of U-based intermetallic magnets is complicated by the dual 
nature of  $5f$ electrons that demonstrate both itinerant and localized character, see, {\it e.g.},
 \cite{Sechovsky98,Fulde06}.
Magnetic moments are thought to be well localized in UGe$_2$, which has a high Curie temperature $T_C\sim 52$~K 
and large ordered moments $m_0 \sim 1.5\mu_B$ per U atom. UCoGe with $T_C\sim 2.4$~K and $m_0 \sim 0.06\mu_B$
is considered as the most itinerant among three materials. An intermediate situation is found for URhGe, which has $T_C=9.5$--9.7~K \cite{Aoki01,Sakarya03} and $m_0 =0.41 \mu_B$  \cite{Levy05}.
The Shubnikov-de Haas  \cite{Yelland11}, the Hall conductivity \cite{Aoki14},
 and the thermoelectric power \cite{Gourgout16} measurements indicate that a Fermi surface reconstruction takes place 
in URhGe close to $H_m$. Based on this observation, an interpretation of $H_m$ as a field-induced Lifshitz transition was 
made in several  studies \cite{Yelland11,Aoki14,Gourgout16,Sherkunov18},
 though no consistent explanation of the magnetic properties was obtained within the itinerant scenario.
Note that the reduced value of U moments in URhGe is, at least, partly related to antiparallel locking
of orbital and spin moments of $5f$ electrons due to the strong spin-orbit coupling \cite{Miiller09,Wilhelm17}.

% = = = = = = = = = = = = = = = = = = = = = = = = = = = = = = = = = = = = = = = = = = = = = = = = = = = 
\begin{figure}[tb]
\centering
\includegraphics[width=0.9\columnwidth]{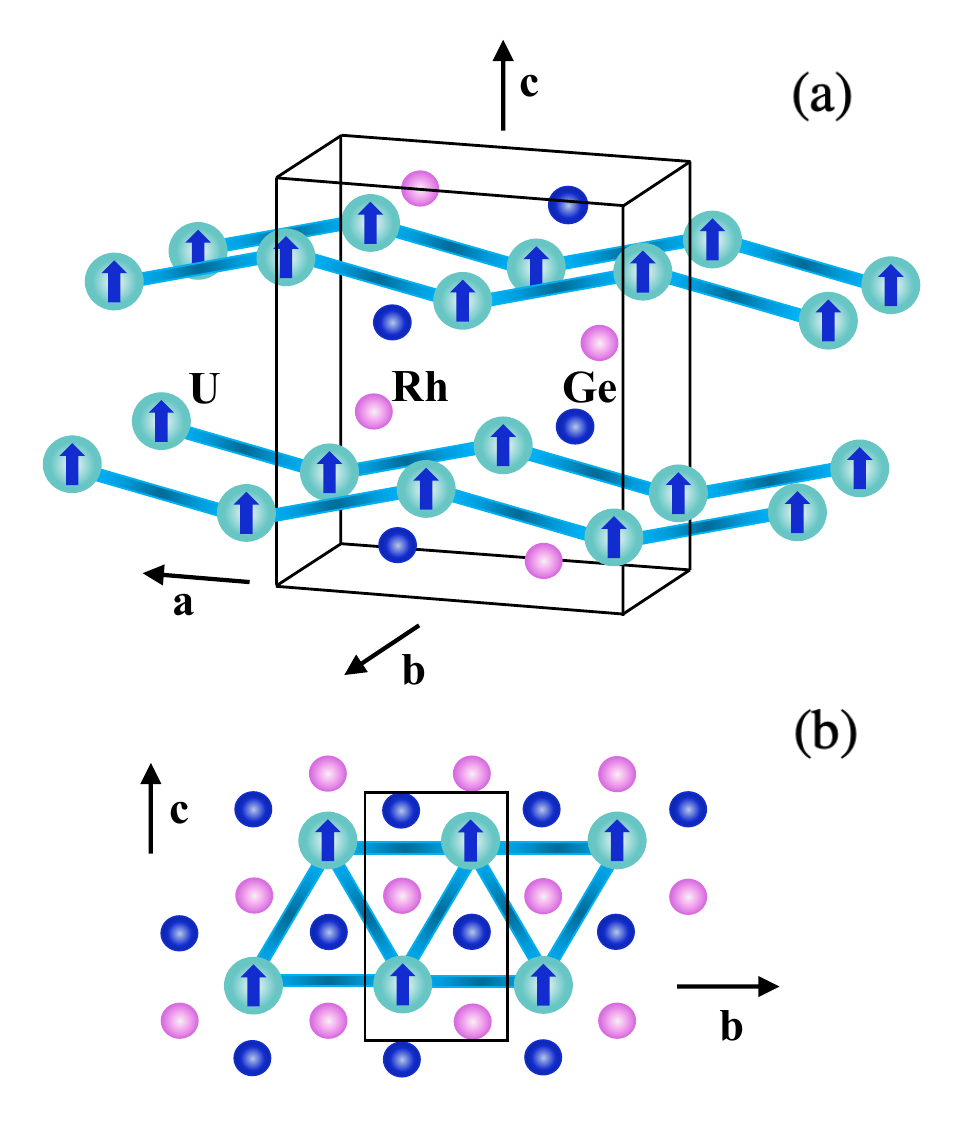}
\caption{Crystal lattice of URhGe. (a) General view, the rectangular prism shows the unit cell. (b) Projection along the $a$ axis.
Only the U atoms with coordinates $x\approx 0$ are included, the shown Rh and Ge atoms are positioned near to 
the $x = 0.2$ plane.}
 \label{fig:crystal}
\end{figure}
% = = = = = = = = = = = = = = = = = = = = = = = = = = = = = = = = = = = = = = = = = = = = = = = = = = = 

In this work, we adopt a local-moment description of the magnetic subsystem in URhGe. Our approach is based 
on the following experimental facts that are hardly consistent with the itinerant picture. First, the magnetic 
susceptibility along three principal directions follows the Curie law  in a wide range of temperatures below 
300~K \cite{Prokes02}. Second, the net in-plane magnetization $\sqrt{m_b^2+m_c^2}$ stays almost constant 
through the metamagnetic transition $H_m$ \cite{Levy05}. 
Furthermore, the X-ray magnetic circular dichroism measurements indicate that $\sim 80$--90\%\ of the total magnetization 
of URhGe is determined by the $5f$ electrons localized on the U atoms \cite{Wilhelm17}. 
A weak, nearly  isotropic growth  of the magnetization observed  in strong magnetic fields $H\parallel bc$  \cite{Levy05} 
can be attributed to the residual contribution of the conduction bands.

URhGe has the orthorhombic TiNiSi-type crystal structure corresponding to the $Pnma$ space group
\cite{Sechovsky98,Prokes02}. The lattice is formed by zigzag chains of U atoms that propagate along 
the $a$ crystallographic direction, Fig.~\ref{fig:crystal}.  The TiNiSi lattice is a derivative of 
the high-symmetry hexagonal AlB$_2$-type structure \cite{Yoshii06}. The connection becomes  transparent 
for the projection along the $a$ direction, which is parallel to the six-fold axis of the hexagonal lattice,  see
Fig.~\ref{fig:crystal}(b). Accordingly, there is a large disparity in magnetic properties of URhGe 
between the $a$ axis and two orthogonal directions. Indeed, at temperatures close to $T_C$, the magnetic 
susceptibility $\chi_a$  is an order of magnitude smaller than $\chi_{b,c}$. The difference between 
$\chi_b$ and $\chi_c$ is further determined by a weak in-plane anisotropy.
 
The abrupt orientational transition in URhGe is a rare case among easy-axis ferromagnets. Usually,
magnetic moments rotate continuously in an applied field until a full alignment is reached at
the second-order transition field. Such 
a behavior is highlighted by the transverse-field Ising model, which often serves as a paradigmatic 
example of the $Z_2$ quantum critical point  \cite{Sachdev}. 
Nonetheless, the first-order transition in a transverse magnetic field can be induced by higher-order harmonics
in the angular dependence of the magnetocrystalline anisotropy \cite{Asti}.
Such a mechanism has been used to explain the first-order magnteization processes
in various ferromagnetic alloys, see, {\it e.g.}, \cite{Melville76}.

For an orthorhombic ferromagnet the energy density as a function of angle $\theta$ measured 
from the easy axis towards an intermediate one is expanded as
\begin{equation}
{\cal E} = K_1\sin^2\theta + K_2 \sin^4\theta  + \ldots
\label{MCA}
\end{equation}
with $K_1>0$.
If only the two leading terms are kept in  (\ref{MCA}), the first-order transition in magnetic field
can appear for negative  $K_2$, see \cite{Asti} and Sec.~III.

The  LSDA calculations of the magnetic anisotropy for URhGe  show a good agreement with Eq.~(\ref{MCA}) 
predicting  a fairly large value of the second harmonic: $K_2/K_1\approx -0.66$ \cite{Shick02}. 
A possible link between the complex form of magnetic anisotropy and the first-order reorientation process 
has been suggested in \cite{Mineev06}, though no explicit calculations were provided.
A phenomenological description of URhGe based on the
Landau expansion for an orthorhombic ferromagnet was formulated in
several works \cite{Levy05,Mineev21,Huxley}. 
However, such a description is not entirely consistent.
The metamagnetic transition in URhGe takes place
at low temperatures and in high magnetic fields, which are beyond the scope of the Landau theory.

Below, we follow an alternative approach and formulate a simple microscopic model,
 which is suitable to study both magnetic statics and  dynamics
of URhGe. In this article, we focus on the static properties and, in particular, explain a high sensitivity of 
the metamagnetic transition to the applied field orientation \cite{Levy07,Levy09} 
as well as obtain location of the tricritical point on the transition line $H_m(T)$  \cite{Gourgout16,Nakamura17}. 
The paper is organized as follows. Section II outlines a spin model chosen  for URhGe.
A zero-temperature analysis of the model is presented in Sec.~III.
Section~IV describes the finite-temperature properties, the tricritical point, and
the $H$--$T$  diagram obtained  with the help of the classical Monte Carlo simulations. 
The obtained results are summarized in Sec.~V. 
The supplemental zero-field Monte Carlo data are included in Appendix.

%%%%%%%%%%%%%%%%%%%%%%%%%%%%%%%%%%%%%%%%%%%%%%%%%%%
\section{Microscopic spin model}
%%%%%%%%%%%%%%%%%%%%%%%%%%%%%%%%%%%%%%%%%%%%%%%%%%%

Quantum effects play a little role in three-dimensional ferromagnets. Therefore, we use a classical spin model
 with unit length spins $\boldsymbol{S}_i$ representing uranium moments.  Our consideration is based on 
the following spin Hamiltonian 
\begin{equation}
\hat{\cal H}  =  - \sum_{\langle ij \rangle}J_{ij}\boldsymbol{S}_i \cdot \boldsymbol{S}_j 
  +  \hat{\cal H}_a   -  \boldsymbol{H} \cdot \sum_i \boldsymbol{S}_i  \ .
\label{H} 
\end{equation}
The first term corresponds to the exchange interactions between uranium moments.
Magnetic anisotropy is accounted for by the single ion term $\hat{\cal H}_a$, although anisotropic exchange interactions 
may also be present in uranium intermetallics. Finally, the Zeeman energy is taken in a simplified form by 
absorbing  an anisotropic $g$ factor and $\mu_B$ into a magnetic field value $H$. 

The single-ion energy is taken as
\begin{equation}
\hat{\cal H}_a = \sum_i \Bigl[DS_i^{x2} + E\bigl( S_i^{y2} - S_i^{z2}\bigr)  + K (S_i^y)^2\,(S_i^z)^2\Bigr],
\label{Ha}
\end{equation}
where  $x,y,z$ are chosen along $a,b,c$, respectively. The first two terms in (\ref{Ha}) is a standard bi-axial anisotropy 
compatible with the point symmetry group on magnetic sites \cite{Buschow}.
The hard  and the easy magnetization directions along the $a$  and the $c$ crystallographic axes correspond to $D\gg E>0$, in accordance with the susceptibility measurements \cite{Braithwaite18} and the LSDA calculations \cite{Shick02}. 
The last term in Eq.~(\ref{Ha}) generates the four-fold $\theta$-harmonic in the macroscopic energy density (\ref{MCA}).
The choice of a microscopic interaction responsible for the $\sin^4\theta$ harmonic is to some
extent  arbitrary. Any appropriate quartic combination of the two spin components
can be equivalently  chosen for $\hat{\cal H}_a$. Also, anisotropic biquadratic interactions can be at the origin of such anisotropy 
and their effect on the thermodynamics will be indistinguishable from the $K$-term  as long as the system has extended
ferromagnetic correlations.

The local symmetry on U sites in the crystal lattice of URhGe consists of the mirror reflection 
$\sigma_y$ only.  Hence, $\hat{\cal H}_a$ may also include a term
$\pm (S_i^xS_i^z + S_i^zS_i^x)$ alternating in sign between four U atoms in the  unit cell 
Fig.~\ref{fig:crystal}. As a result, a uniform ferromagnetic alignment of spins parallel to the $z$  direction can
be accompanied by staggered spin components along $x$. 
The neutron diffraction experiments do not  detect such spin staggering in the ordered state  \cite{Aoki01,Prokes03}.
Hence, we omit  the corresponding term in Eq.~(\ref{Ha}).

Absence of the staggered term  in $\hat{\cal H}_a$ makes four U atoms in a unit cell equivalent and allows 
us to map the real crystal structure onto an orthorhombic Bravais lattice. We assume
exchange interactions  between the nearest neighbors only. 
Furthermore, exchange constants along three crystal directions are replaced by an averaged
exchange parameter $J = (1/z) \sum_j J_{ij}$. The Monte Carlo results included in
Appendix show that the transition temperature of an orthorhombic Heisenberg ferromagnet  depends only weakly
on an  `orthorhombic distortion' of the exchange constants.
Thus, a simplified modelling of an (unknown) three-dimensional exchange pattern in URhGe with a single parameter $J$ does not
introduce a significant quantitative error.

%%%%%%%%%%%%%%%%%%%%%%%%%%%%%%%%%%%%%%%%%%%%%%%%%%%
\section{Zero Temperature}
%%%%%%%%%%%%%%%%%%%%%%%%%%%%%%%%%%%%%%%%%%%%%%%%%%%

In this section we consider the magnetization process in an external field applied parallel to the crystallographic $bc$-plane. At 
zero temperature, spins are confined to the $yz$-plane and remain to be parallel to each other.  Dropping an unimportant constant from the total energy ${\cal E}$, we write it as a function of an angle $\theta$ between the net magnetization 
$\boldsymbol{M}$ and the $z$ axis:
\begin{equation}
{\cal E}/N = (2E + K)\sin^2{\theta} -  K \sin^4{\theta} - H\sin{(\theta+\alpha)}  \,.
\label{eq:E}
\end{equation}
Here $\alpha$ is an angle between an external field and the $y$ axis.
Comparison of Eqs.~(\ref{MCA}) with (\ref{eq:E}) relates the macroscopic and microscopic anisotropy parameters by
$K_1\to (2E + K)$ and $K_2\to -K$.

For small $K$, the energy in zero field increases monotonously  from the minimum value at  
$\theta =0$ to the maximum at  $\theta=\pi/2$.
For $K>2E$ ($K_2/K_1<-0.5$), the orthogonal orientation $\theta=\pi/2$ changes to a local minimum,
whereas a maximum shifts to $\sin^2\theta_0 = (2E+K)/2K$.  In such a situation there is clearly a first-order
transition in the transverse magnetic field, where the equilibrium angle $\theta$ jumps from an intermediate value to
$\theta=\pi/2$. In the following subsection we derive an exact condition for the development of a first-order jump,
which extends considerably the range of relevant $K$ values in comparison to $K>2E$.

% - - - - - - - - - - - - - - - - - - - - - - - - - - - - - - - - - - - - - - - - - - - - - - - - - - - - - - - - - - - - - - - - - - - - - - - - 
\subsection{$H$ along the $b$ axis}
% - - - - - - - - - - - - - - - - - - - - - - - - - - - - - - - - - - - - - - - - - - - - - - - - - - - - - - - - - - - - - - - - - - - - - - - - 

% = = = = = = = = = = = = = = = = = = = = = = = = = = = = = = = = = = = = = = = = = = = = = = = = = = = 
\begin{figure}[t]
\centering
\includegraphics[width=0.75\columnwidth]{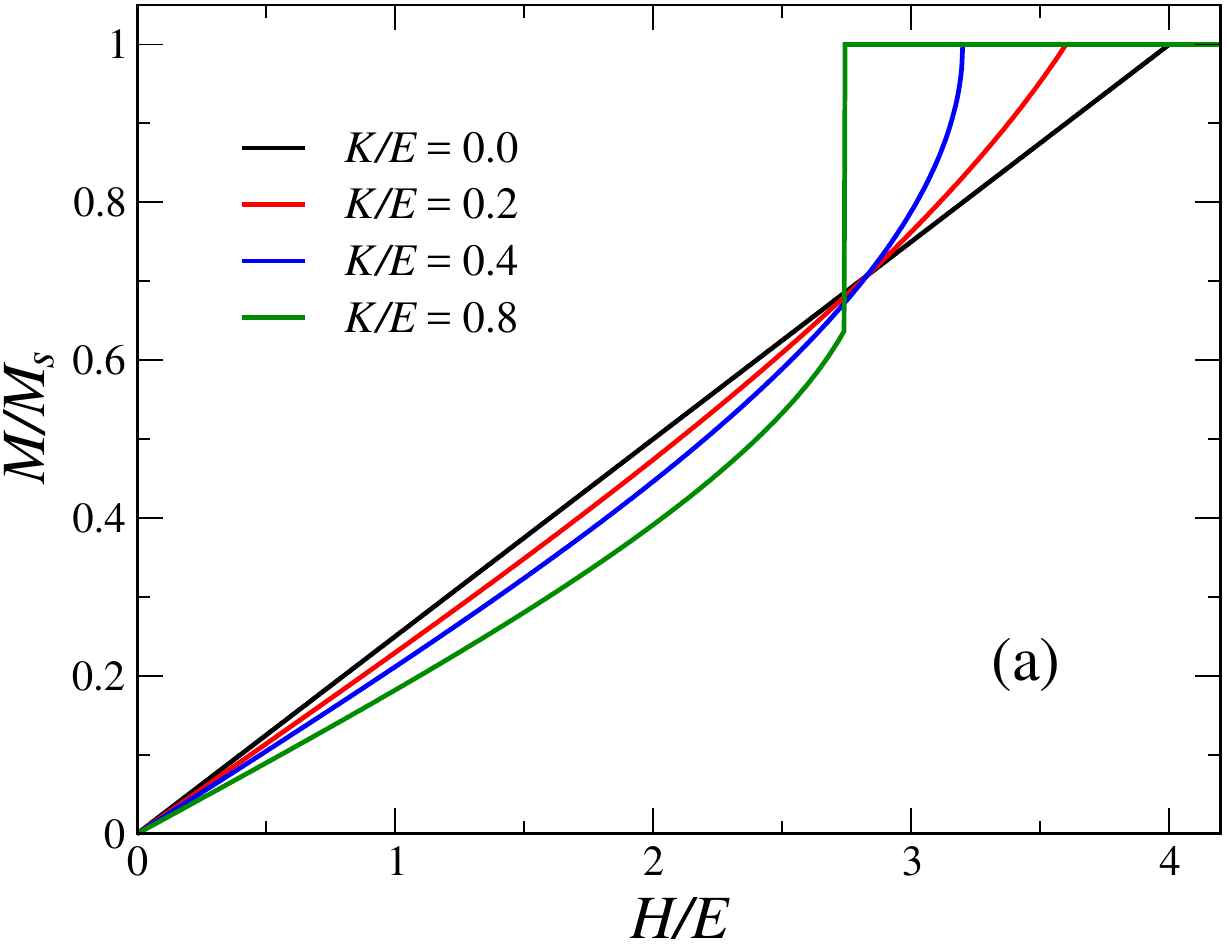}
\vskip 3mm
\includegraphics[width=0.75\columnwidth]{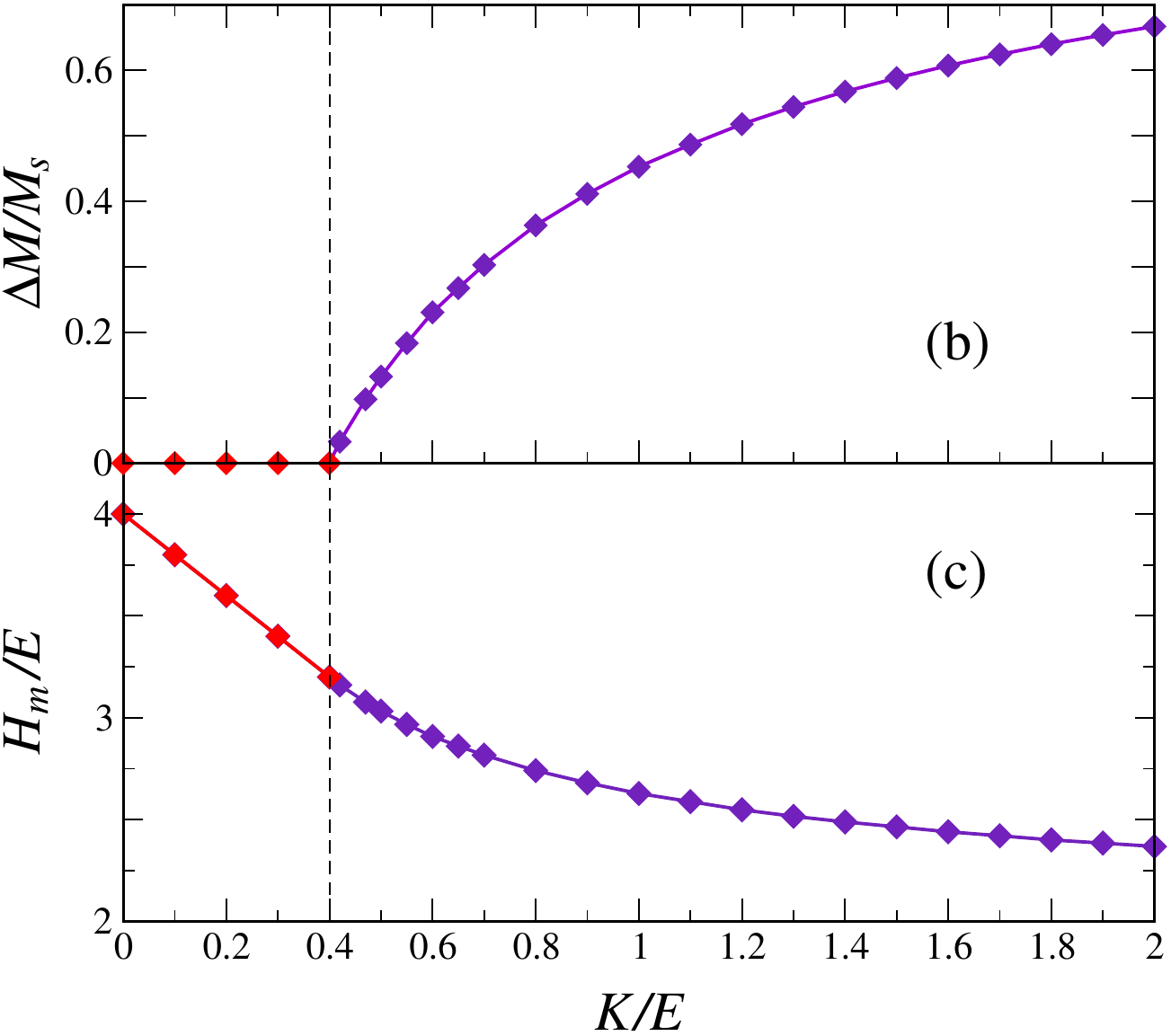}
\caption{(a) Magnetization curves for $H\parallel y$.  Here, $M=M_y$ and $M_s= |\boldsymbol{S}|N$ is the total magnetisation.  (b) The relative magnetization jump and (c) the critical field as a function of $K/E$.}
\label{fig:M_a0}
\end{figure}
% = = = = = = = = = = = = = = = = = = = = = = = = = = = = = = = = = = = = = = = = = = = = = = = = = = = 

The minimum energy condition applied to Eq.~(\ref{eq:E}) yields for $\alpha = 0$: 
\begin{equation}
2(K + 2E)\sin{\theta} - 4 K\sin^3{\theta}   -  H = 0\,.
\label{eq:x3}
\end{equation}
For $K=0$, the magnetization tilts continuously from the easy direction with  $\sin\theta = H/H_m$ until 
a full alignment is reached at the second-order transition field $H_m = 4E$. For finite $K>0$, 
rotation of spins remains continuous  as long as the cubic equation obtained from Eq.~(\ref{eq:x3}):
\begin{equation}
f(x) = ax - bx^3 - H = 0\,, \quad x=\sin\theta \,,
\label{cubic}
\end{equation}
has only a single root for all $H>0$ in the physical domain $0\leq x\leq 1$.
By expanding the energy (\ref{eq:E}) near $x=1$ we obtain that  the second-order 
transition shifts as $H_m = 4E - 2K$ by the extra in-plane term in $\hat{\cal H}_a$.

A second physical root  of Eq.~(\ref{cubic}) appears for certain $H$ once a local maximum of the cubic function at 
$x_{\rm max}^2 = a/3b^2$ shifts from large positive values to $x_{\rm max}<1$. 
This takes place at $2(K + 2E)= 12 K$ or $K/E = 0.4$. Since $f'(x)<0$ for $x>x_{\rm max}$,
the second root at $x_2>x_{\rm max}$ is always a saddle point  of the total energy ${\cal E}(x,H)$ (\ref{eq:E}). 
Hence, if  the second solution  with $0<x_2<1$ is present, 
the magnetization cannot rotate continuously all the way between $\theta =0$ and $\pi/2$ upon increasing $H$
as it would lead to passing through the saddle point. Instead, at certain $H_m$ there is a direct jump into 
the fully aligned state with $\theta =\pi/2$ ($x=1$). Interestingly, transformation to the first-order magnetization process occurs
for relatively small values of higher-order anisotropy constants: $K > 0.4E$  or $K_2/K_1<-1/6$.

The discussed behavior is illustrated in Fig.~\ref{fig:M_a0}(a), which shows the magnetization 
component along the field computed by  a numerical minimization of the classical energy (\ref{H}) as well as by 
directly solving the cubic equation (\ref{eq:x3}).
For the threshold value $K/E=0.4$ between second- and first-order magnetization processes $M(H)$ exhibits 
a pronounced upward curvature with the asymptotic behavior  $M(H) \simeq \sqrt{H_m-H}$ 
near the saturation.
The magnitude of the magnetization jump $\Delta M$ grows continuously above the threshold. 
Theoretical results for $\Delta M/M_s$, where $M_s$ is the total magnetization, are shown in
Fig.~\ref{fig:M_a0}(b). Figure \ref{fig:M_a0}(c) illustrates a variation  the transition field $H_m$ 
for different  $K/E$ values.

An experimental value of the magnetization jump $\Delta M$ can be used to fix the ratio $K/E$  in our  model. 
The low-temperature magnetization measurements yield $\Delta M/M_s\sim 0.3$ for URhGe   \cite{Nakamura17}.
This jump value corresponds to $K/E \approx 0.7$ and  $H_m \approx 2.82E$, see Fig.~\ref{fig:M_a0}(b,c).  
For the magnetocrystalline anisotropy expansion (\ref{MCA}) one finds accordingly  $K_2\approx -0.26K_1$.
Hence, the experimental data yield a significantly smaller ratio of two anisotropy constants in comparison to 
the LSDA results \cite{Shick02}. From the experimental value of 
the transition field $H_m=11.7$~T and ordered moments $M_s = g^*\mu_B\approx 0.41\mu_B$
 we obtain in the physical units: $E = g^*\mu_B H_m/2.82 \approx 0.098$~meV or 1.14~K.

The uniaxial-stress measurements in magnetic field $H\parallel b$ observe a fast suppression 
of the transition field  $H_m$ for moderate stress $\sigma_b$ 
applied along the $b$ crystal axis \cite{Braithwaite18}. In addition, 
the magnetization slope $dM/dH$ rapidly increases with $\sigma_b$, whereas the Curie temperature $T_C$ 
stays almost constant. Since both quantities $H_m$ and $dH/dM$ are set by the magnitude of
the in-plane anisotropy constant, the experimental results suggest that $E$ is strongly reduced by the 
$b$-axis stress. Such a behavior is consistent 
with a gradual restoration of structural isotropy in the $bc$ plane by removing distortion of Rh-Ge hexagons, 
see Fig.~\ref{fig:crystal}.  By the same token, $K$ goes to zero as well, since the $\sin^4\theta$ harmonic
is also incompatible with the hexagonal rotation symmetry. On the other hand, the Curie temperature  is set by 
exchange interactions between U atoms, which experience much weaker variations for $\sigma_b\alt 0.5$~GPa.

% - - - - - - - - - - - - - - - - - - - - - - - - - - - - - - - - - - - - - - - - - - - - - - - - - - - - - - - - - - - - - - - - - - - - - - - -
\subsection{Tilted magnetic field}
% - - - - - - - - - - - - - - - - - - - - - - - - - - - - - - - - - - - - - - - - - - - - - - - - - - - - - - - - - - - - - - - - - - - - - - - -
 
The second-order Ising transition  is smeared  once an external field rotates toward the easy axis. 
Still, the first-order metamagnetic  transition remains stable for a range of 
tilting angles. The magnetization jump
is continuously reduced and vanishes at a certain angle $\alpha^*$. A high sensitivity of the reorientation transition
$H_m$ in URhGe to the magnetic field direction has been reported by a number of authors \cite{Levy09,Nakamura17}.

% = = = = = = = = = = = = = = = = = = = = = = = = = = = = = = = = = = = = = = = = = = = = = = = = = = = 
\begin{figure}[tb]
\centering
\includegraphics[width=0.75\columnwidth]{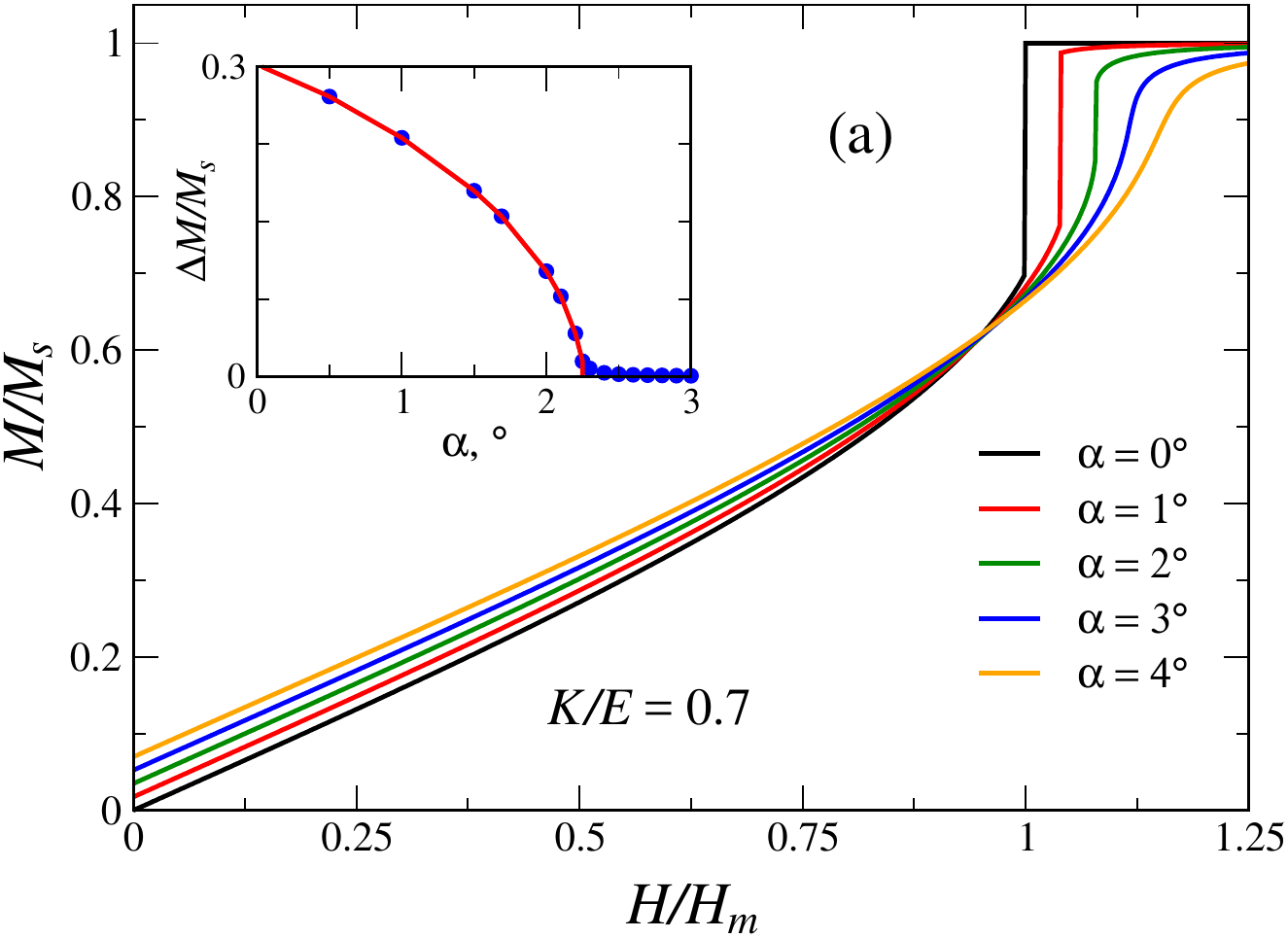} 
\vskip 3mm
\includegraphics[width=0.72\columnwidth]{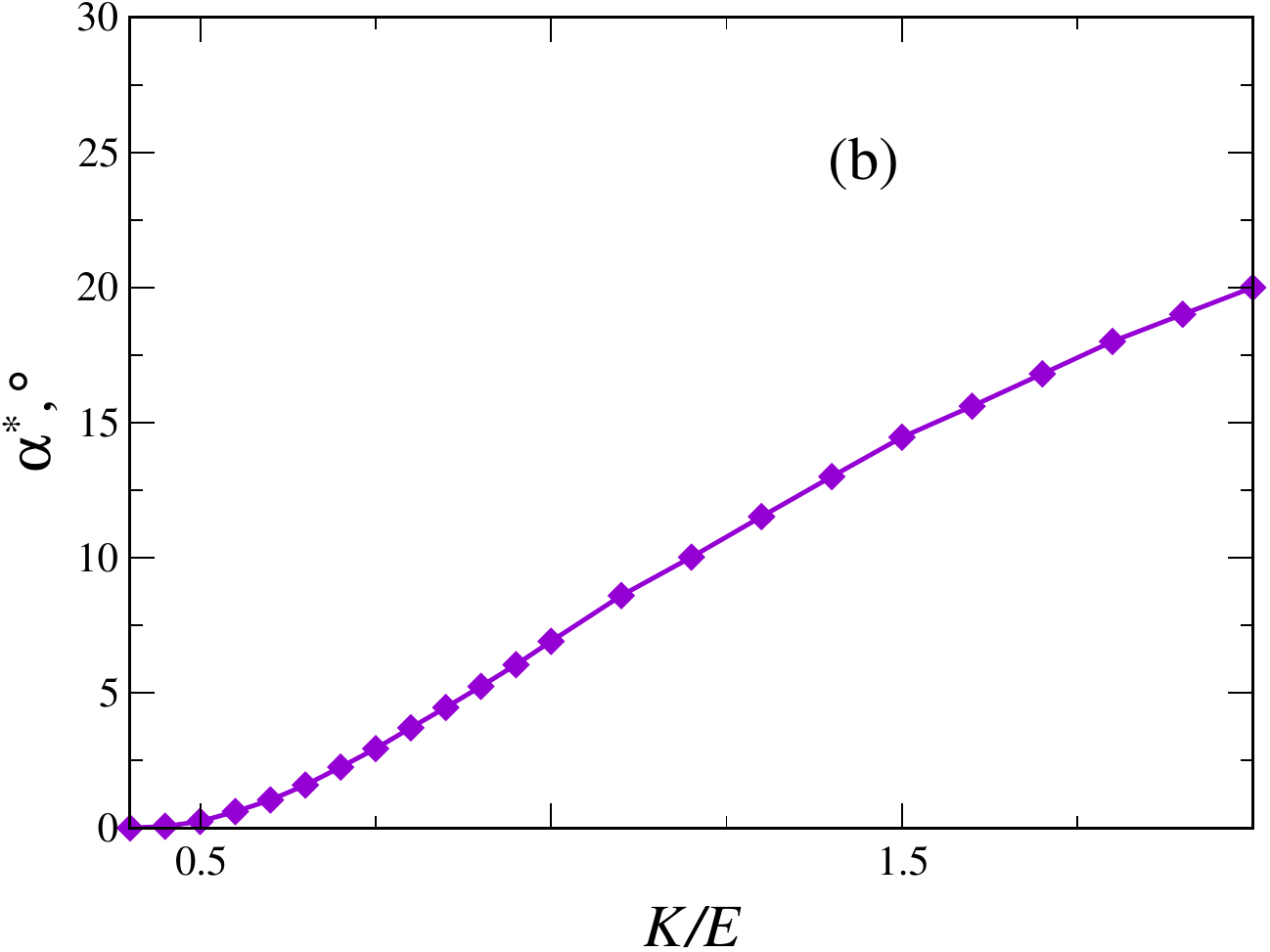}
\caption{(a) Magnetization curves for finite tilting angles $\alpha$ between an applied field and the $b$ axis, 
$K=0.7E$.  The inset shows the tilting-angle dependence of the magnetization jump  $\Delta M(\alpha)$ 
used to determine the critical angle $\alpha^*$.
(b) Dependence of the critical angle on the  anisotropy constant.}
\label{fig:KEalpha}
\end{figure}
% = = = = = = = = = = = = = = = = = = = = = = = = = = = = = = = = = = = = = = = = = = = = = = = = = = = 

Figure \ref{fig:KEalpha}(a) shows the longitudinal magnetization  $M_H = (\boldsymbol{M}\cdot{\boldsymbol H})/H$
in a tilted field $\tan\alpha = H_z/H_y$ obtained by numerical minimization of the energy (\ref{eq:E}) for  $K/E=0.7$.
As the field tilting progresses, a magnitude of the magnetization jump  is quickly suppressed. 
The jump vanishes at a critical point $(\alpha^*,H_m^*)$ on the first-order transition line $H_m(\alpha)$.
By extrapolating $\Delta M(\alpha)$ to zero, we find the critical angle value $\alpha^* = 2.25^{\circ}$, 
see the inset in Fig.~\ref{fig:KEalpha}(a). The corresponding magnetic field  is $H_m^*\approx 1.1H_m$ or 
$12.9$~T in the dimensional units. Figure \ref{fig:KEalpha}(b) shows the dependence of 
critical angle $\alpha^*$ on $K/E$.

The above theoretical  values for $\alpha^*$, $H_m^*$ calculated with $K/E=0.7$ are somewhat smaller than 
the experimental results $\alpha^* \approx 5^\circ$ and  $H_m^*=13.5$~T reported
in Ref.~\cite{Nakamura17}.  The difference may be attributed to an unknown contribution of band
electrons into $M_s$, which increases a relative value of the magnetization jump to be used
in the spin model. Instead, $K/E$ can be estimated from
the experimental $\alpha^*$, although measurements of the critical angle are not  very precise.
In addition, a higher order harmonic $K_3 \sin^6\theta$ in 
the magnetic anisotropy (\ref{MCA}) can play a role for URhGe, especially, under the uniaxial $\sigma_b$ 
stress, which reduces the orthorhombic anisotropy, {\it i.e.}, $K_1$ and $K_2$ ($E$ and $K$). 
Therefore, an inclusion of the corresponding term into ${\cal H}_a$ may be necessary 
for a better fit of the experimental data. Since our aim in this work is to introduce a basic theoretical framework, 
we still consider the minimal spin model (\ref{H}) with $K/E=0.7$.

%%%%%%%%%%%%%%%%%%%%%%%%%%%%%%%%%%%%%%%%%%%%%%%%%%%
\section{Finite Temperatures}
%%%%%%%%%%%%%%%%%%%%%%%%%%%%%%%%%%%%%%%%%%%%%%%%%%%

We now turn to the finite-temperature properties of the spin model (\ref{H}), which have been studied using
the classical Monte Carlo simulations. The standard Metropolis algorithm was combined with 
a restricted motion of spins in order to keep an acceptance rate at the level of 40--50\%. Specifically, 
a trial spin orientation  is randomly chosen on a spherical cap rather than on the whole sphere. 
The cap is centered on the initial spin direction and its height depends on temperature according to 
$\Delta S^z \simeq T$. In the present work  we set the first $10^5$ Monte Carlo steps 
at each temperature/field point for thermal equilibration and performed measurements over subsequent
$5 \cdot 10^5$  steps. The Monte Carlo results were additionally averaged over 50--200 independent runs initialized
by different random spin configurations. Such a procedure also provides an unbiased estimate of the statistical 
errors.

The analysis of Sec.~IIIA allows us to fix the absolute values of 
the in-plane anisotropy constants $E$ and $K$. In addition, the exchange parameter $J$ 
can be inferred from the measured Curie temperature $T_C=9.7$~K \cite{Sakarya03}. 
This step is complicated by the fact that magnetic anisotropy also affects the transition temperature.
We have adopted the  following  procedure.
URhGe has a dominant planar
anisotropy, which places it in between the Heisenberg and the $XY$ ferromagnets. 
For the latter two models, transition temperatures are, respectively,  $T_c=1.4429 J$ \cite{Chen93} and 
$T_c =2.2016J$ \cite{Hasenbusch90}.  Bracketing $T_c$ for the spin model (\ref{H})
between these two values and using $E \approx 1.14$ K we obtain a relevant interval for 
$E/J\in (0.17,0.3)$.  A rough estimate $D/J=3$ was also made based on the anisotropy of the magnetic susceptibility 
in the paramagnetic regime \cite{Braithwaite18}. After that, a series of trial Monte Carlo runs
was performed for various sets of microscopic constants in the chosen interval with the aim to fit 
the Curie temperature for URhGe. The obtained best parameter values are
 $E/J = 0.22$ and $J=5.18$~K. A small value of $E/J$ shows that URhGe
is far from being an Ising ferromagnet as commonly assumed in the literature. Instead it has to
be described as an $XY$ ferromagnet with  a weak in-plane anisotropy.

\subsection{Zero magnetic field}

 % = = = = = = = = = = = = = = = = = = = = = = = = = = = = = = = = = = = = = = = = = = = = = = = = = = = 
\begin{figure}[tb]
\centering
\includegraphics[width=0.77\columnwidth]{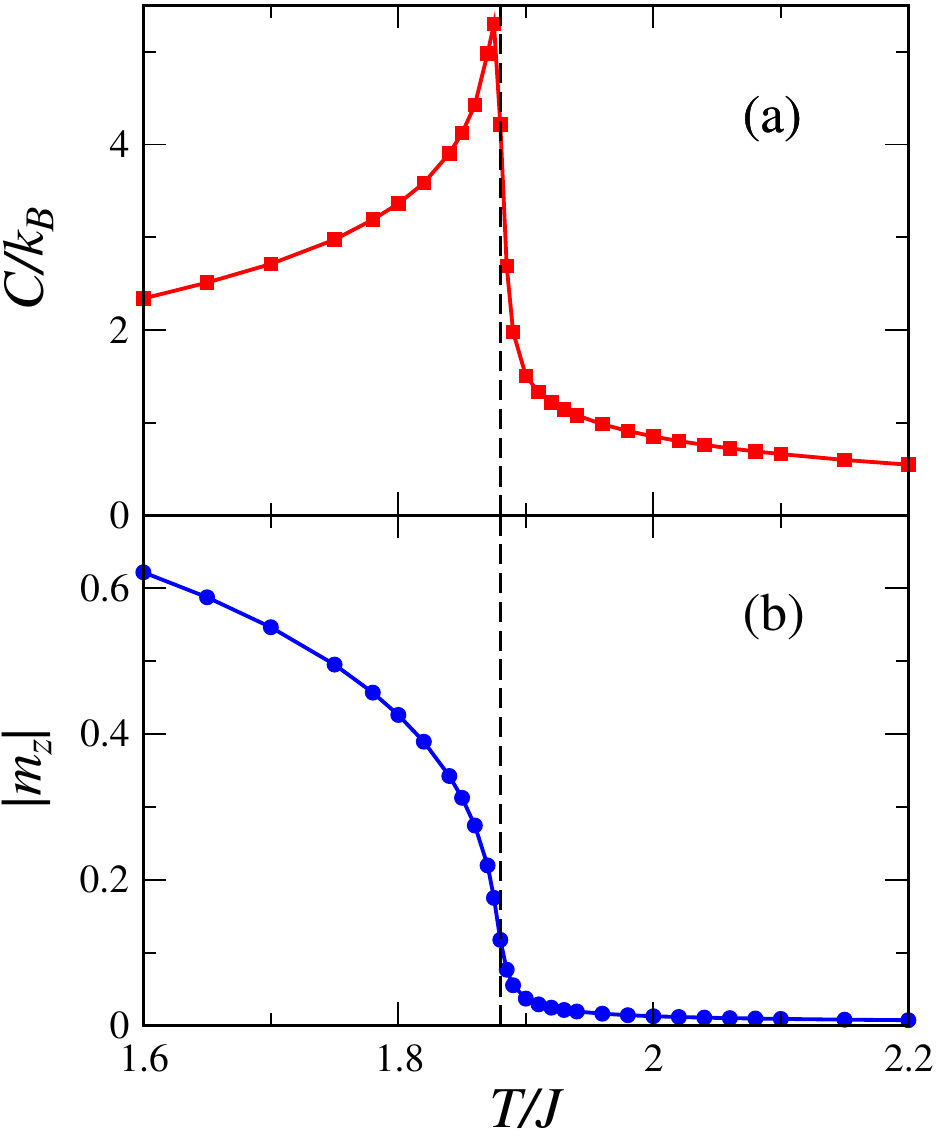} 
\caption{Temperature dependence of (a) the specific heat and (b) the ferromagnetic
magnetization for the spin model of URhGe with 
$E/J= 0.22$, $D/J=3$, and $K/E =0.7$ obtained for a spin cluster with $L=36$.
The vertical dashed line shows the position of the transition point at $T_c/J = 1.880$ 
in the thermodynamic limit $L\to\infty$}.
\label{fig:CMvsT}
\end{figure}
% = = = = = = = = = = = = = = = = = = = = = = = = = = = = = = = = = = = = = = = = = = = = = = = = = = = 

The Monte Carlo simulations for the selected set of microscopic parameters have been performed on 
cubic clusters with $N=L^3$ spins and linear sizes $L=8$--40. Figure \ref{fig:CMvsT} shows 
the temperature dependence of
the heat capacity $C$ and the spontaneous 
magnetization $\langle |m_z|\rangle$  close to a phase transition. 
The $\lambda$-like anomaly in $C(T)$  gives a clear indication of the second-order phase transition
at $T_c/J \sim 1.86$. 
A  continuous rise of  the ferromagnetic order parameter $m_z$ below $T_c$ further supports this conclusion. 
Still, the finite-size effects smear sharp
singularities in the physical quantities and produce a rounding-off behavior near $T_c$ for both $C(T)$ and $m_z(T)$.
The specific heat peak is also slightly displaced away from the bulk $T_c$, see  Fig.~\ref{fig:CMvsT}.

\begin{figure}[tb]
\centering
\includegraphics[width=0.77\columnwidth]{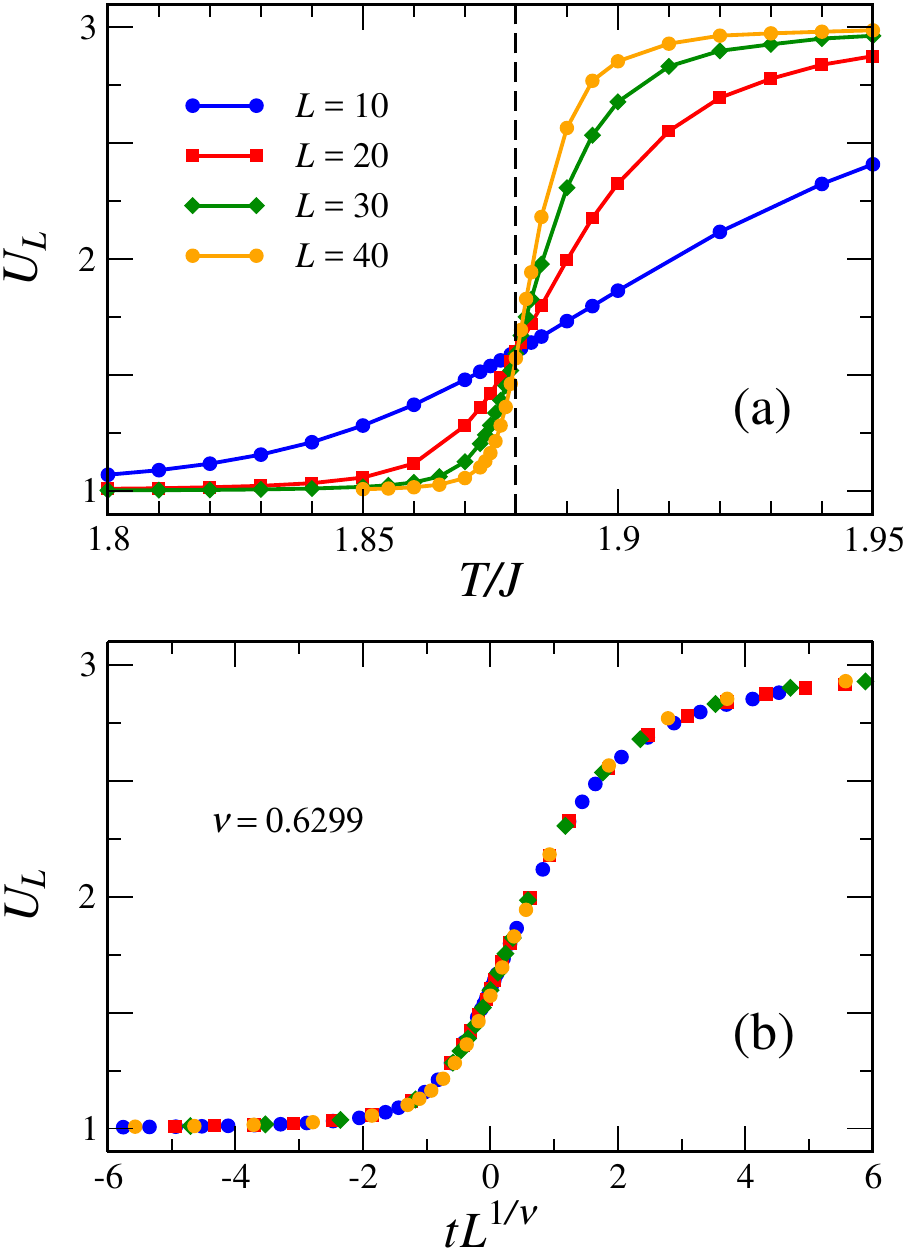} 
\caption{The fourth-order cumulants $U_L(T)$ for different lattices
as a function of (a) temperature and (b) the scaling variable $t L^{1/\nu}$.
Same symbols are used to represent lattices with the same linear sizes $L$ on both panels.
}
\label{fig:UvsT}
\end{figure}

A precise location of the transition temperature $T_c$ can be obtained using the
fourth-order cumulant approach \cite{Binder81,Binder}. The cumulants defined by
\begin{equation}
U_L = \frac{\langle m_z^4\rangle}{\langle m_z^2\rangle^2}
\end{equation}
are computed for lattice clusters of different linear sizes $L$ and plotted as a function of temperature,
see Fig.~\ref{fig:UvsT}(a). In the paramagnetic phase $T\gg T_c$ the fluctuations of the order
parameter are gaussian  and $U_L\approx 3$. Below $T_c$, the order parameter
acquires a constant value and $U_L\approx 1$. Near the second-order transition, the finite-size scaling hypothesis 
\cite{Barber83} predicts that 
\begin{equation}
U_L = \tilde{f}(L/\xi) = f (t L^{1/\nu}) \,,
\label{Uscale}
\end{equation}
where $t = (T-T_c)/T_c$, 
$\xi \sim t^{-\nu}$ is a bulk correlation length, and $f(x)$ is a universal scaling function. 
According to the scaling law (\ref{Uscale}),  the cumulant curves $U_L(T)$  cross
 at the critical temperature $t=0$ of an infinite system $L\to\infty$
  \cite{Binder81}. Additional small finite-size
 corrections to the leading scaling behavior can be taken into account by a proper extrapolation
 of the crossing points \cite{Chen93}.
     
 % = = = = = = = = = = = = = = = = = = = = = = = = = = = = = = = = = = = = = = = = = = = = = = = = = = = 
% = = = = = = = = = = = = = = = = = = = = = = = = = = = = = = = = = = = = = = = = = = = = = = = = = = = 

The  transition temperature obtained from the crossing points of the fourth-order cumulants is $T_c/J = 1.880(1)$,
which amounts  to $T_C = 9.76$~K in excellent agreement with 
the experimental value. We have also checked that despite
a weakness of the in-plane anisotropy,  $E\ll D$, the critical behavior of the spin model (\ref{H}) still belongs
to the 3D Ising universality class. For that we rescale the Monte Carlo data  for  $U_L(T)$ according to the
scaling law (\ref{Uscale}) with the Ising critical exponent $\nu = 0.6299$ \cite{Ferrenberg18}.
As a result, the data from different clusters collapse  on the common curve given by $f(x)$, 
see Fig.~\ref{fig:UvsT}(b). Additional scaling plots for the order parameter and the susceptibility that 
further verify the Ising values for  $\beta=0.3263$ and $\gamma=1.2371$  are included in Appendix.

The criticial properties  of URhGe  have been measured  and discussed alongside with the experimental 
results for some other uranium ferromagnets in \cite{Tateiwa14,Tateiwa19}. 
The experimental values for the order-parameter exponent $\beta$ are generally quite close to the 3D Ising value.
On the other hand,  the derived critical exponents $\gamma$ and $\delta$ appear to be more mean-field like
across the whole series of studied materials. The source of this discrepancy is currently not clear.

 \subsection{Tricritical point and phase diagram}

  % = = = = = = = = = = = = = = = = = = = = = = = = = = = = = = = = = = = = = = = = = = = = = = = = = = = 
\begin{figure}[b]
\centering
\includegraphics[width=0.75\columnwidth]{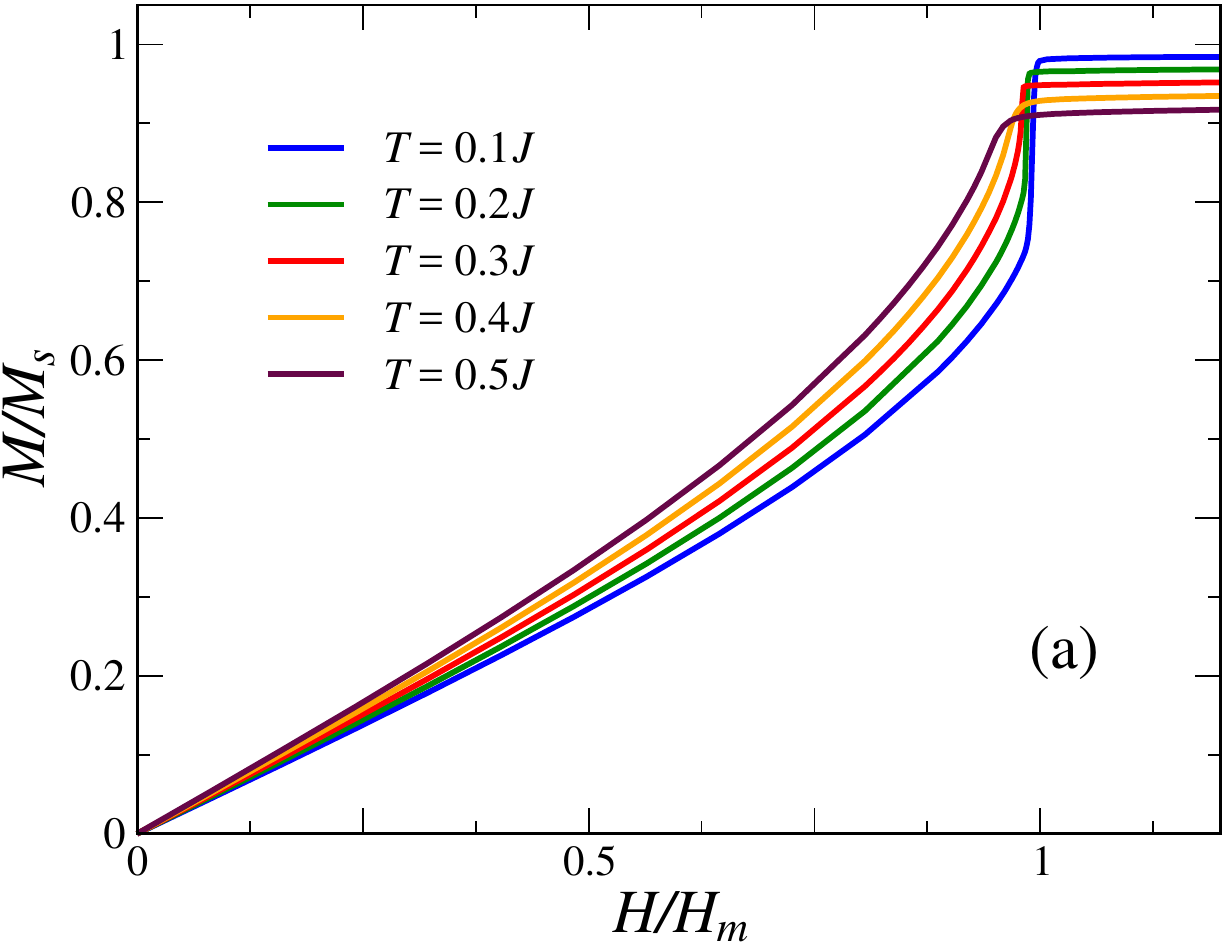}
\vskip 3mm
\includegraphics[width=0.77\columnwidth]{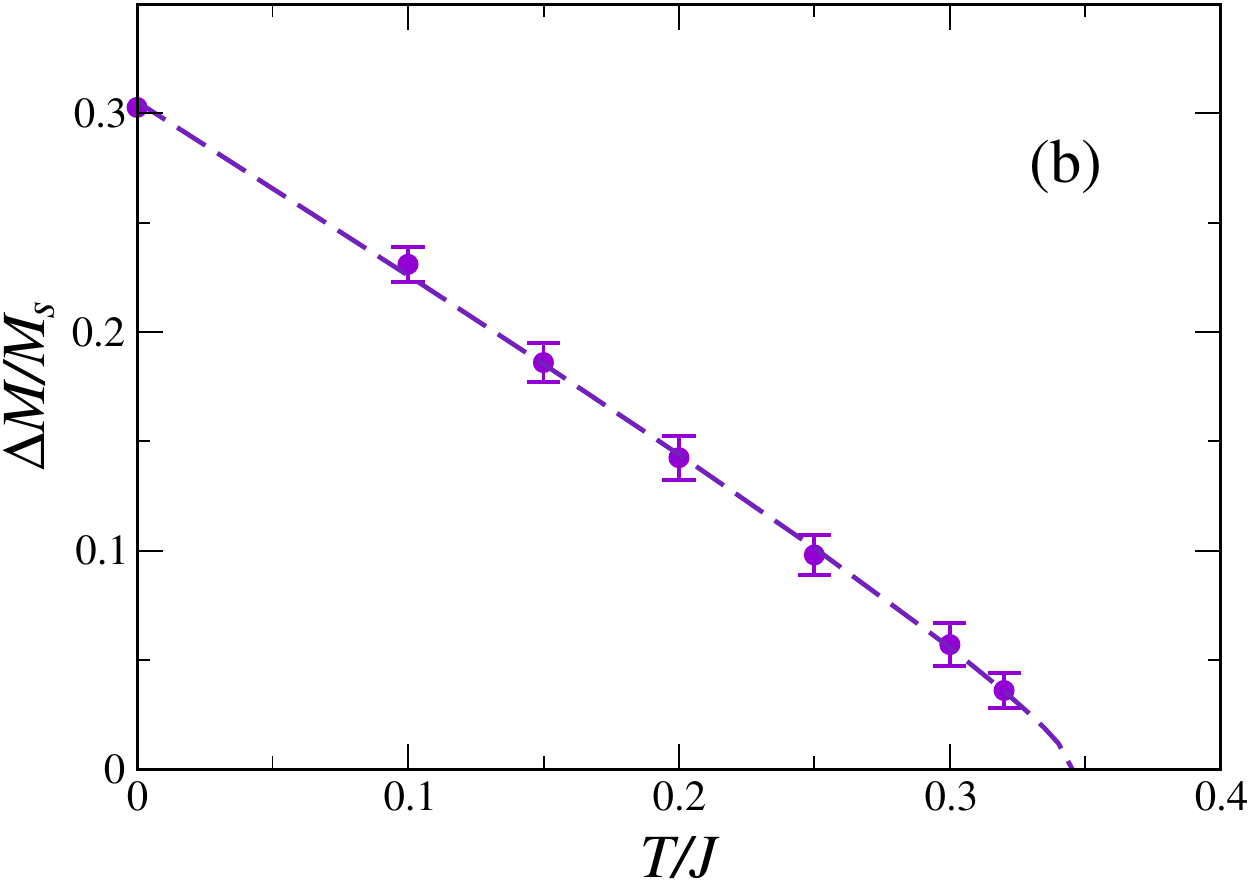}
\caption{(a) Low-temperature magnetization curves. Magnetic field is normalized to 
the value of the first-order transition field at $T=0$: $H_m =0.620J= 2.82E$. 
(b) Variation of the magnetization jump with temperature.
The dashed line is an extrapolation curve used to determine $T_{\rm tr}$.}
\label{fig:mag_curves}
\end{figure}
% = = = = = = = = = = = = = = = = = = = = = = = = = = = = = = = = = = = = = = = = = = = = = = = = = = = 

Let us now consider the behavior in a magnetic field applied parallel to the $b$ axis ($y$ axis). 
Magnetization curves computed in 
the Monte Carlo simulations for a range of temperatures from $T/J=0.1$ to $0.5$ are shown in Fig.~\ref{fig:mag_curves}(a).  The low-temperature curves demonstrate  clear jumps that
signify a first-order transition between the polarized paramagnetic state at $H>H_m$ and the
state with transverse ferromagnetic order at $H<H_m$.
As temperature increases, height of the jump 
goes down and vanishes at a certain temperature. Nature of the phase transition 
changes from the first to the second order at such a tricritical point \cite{LLv5}.
Tricritical points in the phase diagrams of the condensed matter systems have been the subject of 
theoretical and experimental investigations over several decades,
see, for example,
\cite{Griffiths70,Vettier73,Kincaid75,Griffiths73,Wegner73,Stephen75,Fisher78,Stryjewski77,
Gerzanich81,Ziman82,Landau76,Diep87,Herrmann93,Ren06,
Santos10,Belitz99,Kirkpatrick12,Brando16,Taufour16}.
The mean-field Landau theory assigns the tricritical point to a point,
where  the quartic term coefficient pathes through zero \cite{LLv5,Kincaid75}. 
An Ising antiferromagnet  in a longitudinal field provides 
an example of the tricritical point in the $H$--$T$ diagram \cite{Stryjewski77}. 
The tricritical point is also present in the  $p$--$T$ diagram of a metallic
 ferromagnet  with a first-order quantum transition induced by soft 
 fermionic modes   \cite{Belitz99,Kirkpatrick12,Brando16}.

In our Monte Carlo simulations, the tricritical point was located by extrapolating  the magnetization 
jumps to zero $\Delta M\to 0$, as shown in Fig.~\ref{fig:mag_curves}(b). The extrapolation 
yields the tricritical temperature for the chosen set of parameters as $T_{\rm tr} = 0.345(5)J$.
With the previously deduced exchange constant $J=5.18$~K, this corresponds to $T_{\rm tr} = 1.79$~K.
Our theoretical value is in a good agreement with  $T_{\rm tr} \approx 2$~K  reported for URhGe in \cite{Gourgout16}.
Magnetic field at the tricritical point was obtained from the crossing point of the fourth-order cumulants 
$U_L(H) =  \langle m_z^4\rangle/\langle m_z^2\rangle^2$  ($H\parallel y$) computed at $T=T_{\rm tr}$,
similar to the procedure detailed in the preceding subsection for the zero-field transition. In units of the exchange constant
the magnetic field value of the tricritical point is $H_{\rm tr} = 0.607(1)J$. Taking into account that 
$\left.H_m\right|_{T=0} = 2.82 E = 0.62J$, this yields $H_{\rm tr} = 11.4$~T.

Fluctuations near a tricritical point are generally stronger than those near
a conventional second-order transition and are characterized by
a nontrivial set of critical exponents already at the mean-field level \cite{LLv5,Kincaid75}:
\begin{equation}
\alpha = 1/2,\quad \beta = 1/4,\quad \gamma = 1,\quad \nu = 1/2\,.
\end{equation}
The renormalization-group arguments indicate that these exponents remain valid for three-dimensional
systems up to multiplicative logarithmic corrections \cite{Wegner73,Stephen75,Fisher78}.
Note that in our case the `specific heat' exponent $\alpha$ applies to the second-order derivative of the thermodynamic
potential $\partial^2 F/(\partial \zeta)^2 \simeq (\zeta - \zeta_{\rm tr})^{-\alpha}$ along an arbitrary path
$\zeta(T,H)$ in the $T$--$H_y$ plane that crosses the transition line $H_m(T)$ under a finite angle. In particular,
near the tricritical point the field derivative of the magnetization diverges with the mean-field exponent $\alpha$ \cite{Fisher78}:
\begin{equation}
d M/dH =(H - H_{\rm tr})^{-1/2} \,.
\end{equation}
Such a square-root singularity is clearly seen
in the behavior $M(H)$ shown in Fig.~\ref{fig:mag_curves}(a), though we do not attempt here a quantitative comparison.

% = = = = = = = = = = = = = = = = = = = = = = = = = = = = = = = = = = = = = = = = = = = = = = = = = = = 
\begin{figure}[t]
\centering
\includegraphics[width=0.75\columnwidth]{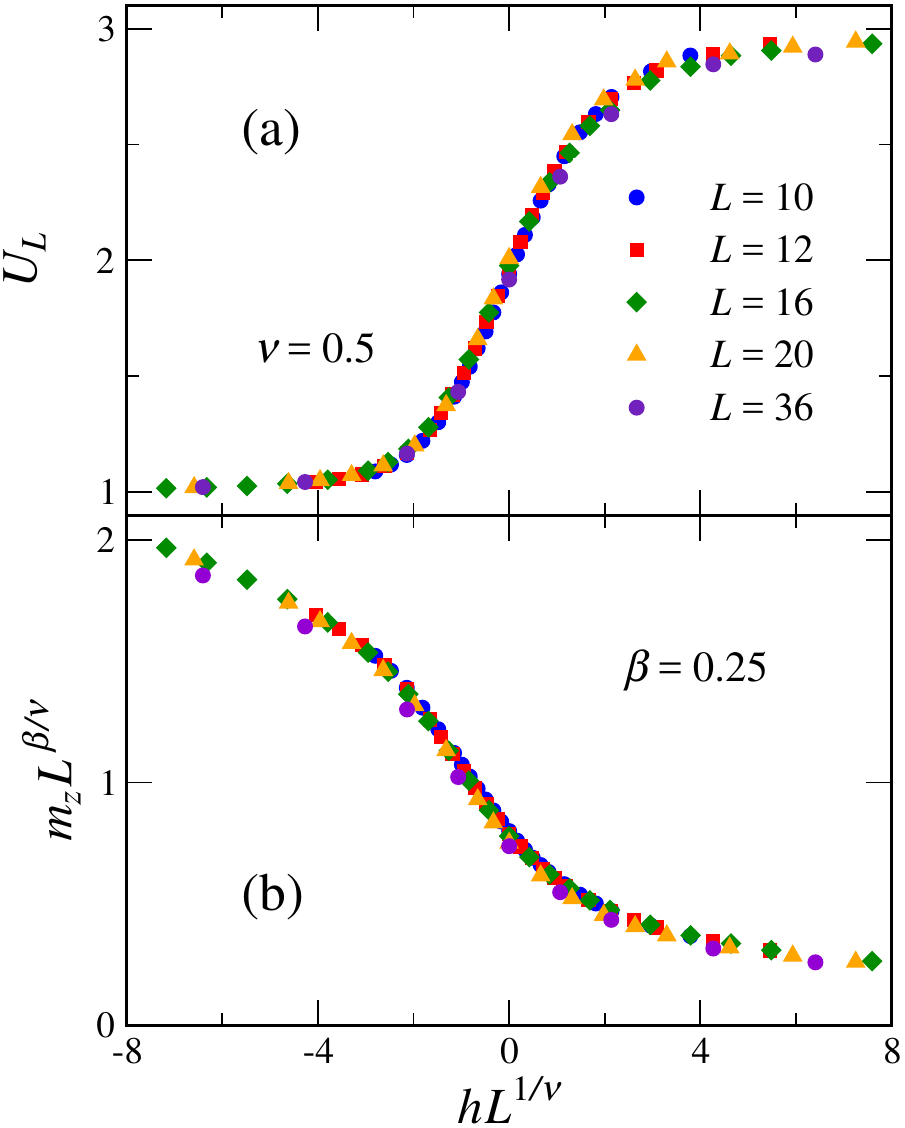}
 \caption{Scaling plot for (a) the fourth-order cumulants $U_L(H)$ and (b) the order parameter $m_z$
 at $T=T_{\rm tr}$. The tricritical mean-field exponents $\nu=1/2$, $\beta=1/4$ are used. 
 }
\label{fig:SPH}
\end{figure}
% = = = = = = = = = = = = = = = = = = = = = = = = = = = = = = = = = = = = = = = = = = = = = = = = = = = 

We further verify the tricritical exponents $\beta$ and $\nu$ 
by scaling the Monte Carlo results for $U_L(H)$ and $m_z(H)$ calculated for clusters with different linear sizes.
The scaling parameter for an  isothermal field scan is $h L^{1/\nu}$, where $h=(H-H_{\rm tr})/H_{\rm tr}$.
The order parameter behavior in the critical region $m_z \simeq (H_{\rm tr}-H)^\beta$,
corresponds to a finite-size scaling form 
\begin{equation}
m_z =L^{-\beta/\nu} g (h L^{1/\nu})\,.
\end{equation}
Accordingly,
the data for $m_z$ need to be compensated by the cluster dependent factor $L^{\beta/\nu}$. The scaling
plots for $U_L(H)$ and $m_z(H)$  are shown in Fig.~\ref{fig:SPH}. 
The data collapse quality is almost as good as for the zero-field transition despite 
the unaccounted logarithmic 
corrections. Overall, the Monte Carlo results of Fig.~\ref{fig:SPH} provide
a firm evidence that the tricritical point is characterized by the mean-field values $\beta=1/4$ and $\nu=1/2$. 
Note that the previous Monte Carlo studies have considered tricritical points for
the Ising spin models only 
\cite{Landau76,Diep87,Herrmann93,Ren06,Santos10}. This work extends 
the numerical analysis of the tricritical behavior to a realistic spin Hamiltonian with three-component 
magnetic moments.

Our  Monte Carlo data do not show any significant enhancement in the field-dependent heat capacity 
$C(H)$ as $H\to H_{\rm tr}$ (or $H_m$). In contrast, a  20-25\%\ rise of the specific heat
 between $H=0$ and $H=H_m$  has been observed  in experiment 
 \cite{Levy09}. One possible explanation for this discrepancy is that the 
observed increase is entirely due to the conduction electrons 
either via  a reconstruction of the Fermi surface
\cite{Yelland11,Aoki14,Gourgout16} or via the effective mass enhancement.

% = = = = = = = = = = = = = = = = = = = = = = = = = = = = = = = = = = = = = = = = = = = = = = = = = = = 
\begin{figure}[t]
\centering
\includegraphics[width=0.77\columnwidth]{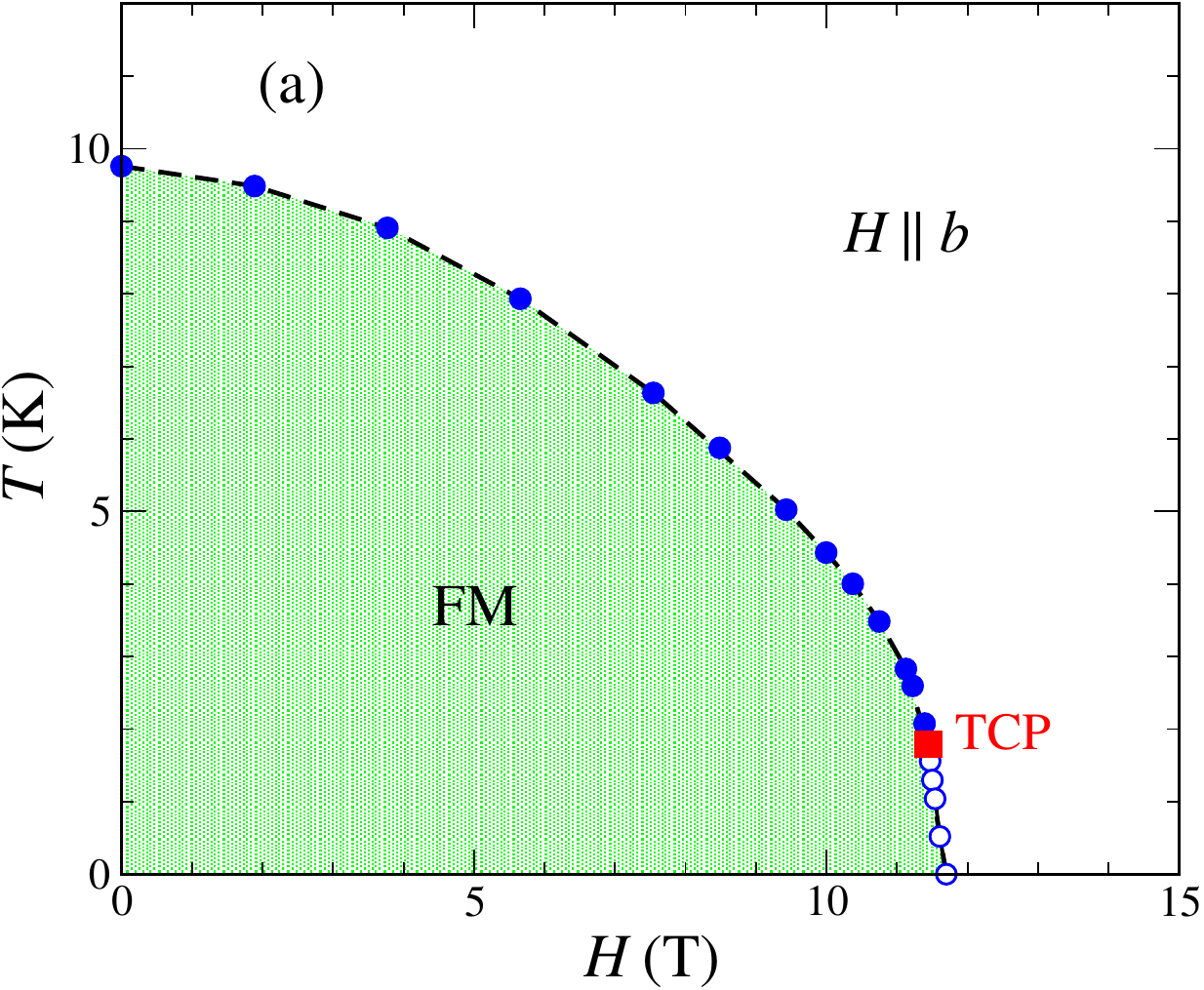} 
\vskip 3mm
\includegraphics[width=0.8\columnwidth]{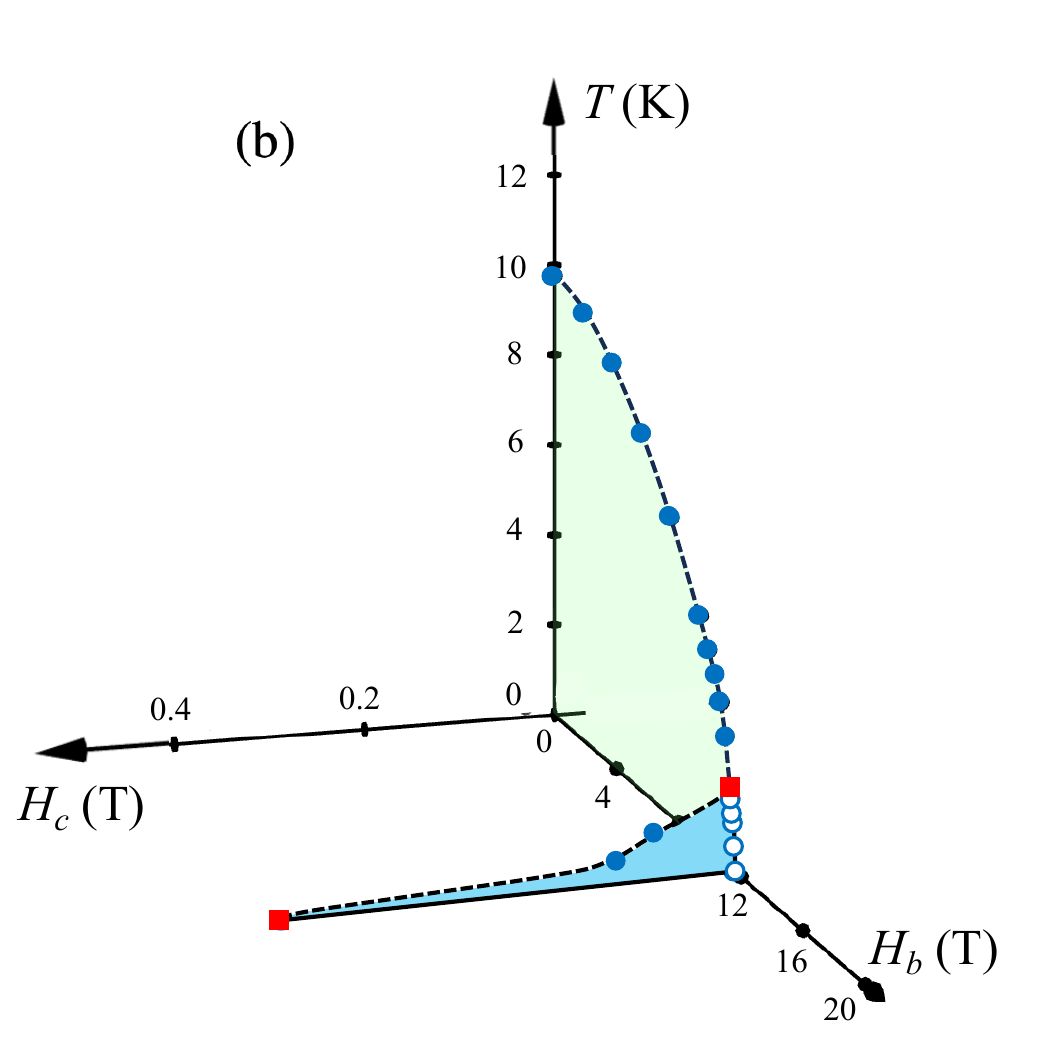}
\caption{(a) The Monte Carlo phase diagram of the anisotropic spin model for URhGe in a 
magnetic field parallel to the $b$ axis. FM is a ferromagnetic phase. 
Closed circles and dashed lines indicate the continuous second-order transitions, 
open circles represent the first-order transitions.
Red square marks the tricritical point (TCP). 
(b) The $T$--$H_b$--$H_c$ phase diagram of URhGe. 
The blue surface corresponds to the plane of first-order transitions 
bounded at finite temperatures by a critical line $\alpha^*(T)$ 
that starts at TCP and extends down to $T=0$ (second red square).  
}
\label{fig:pd}
\end{figure}
% = = = = = = = = = = = = = = = = = = = = = = = = = = = = = = = = = = = = = = = = = = = = = = = = = = = 

After determining the location of the tricritical point, we
ran the Monte Carlo simulations over a wide range of temperatures 
and magnetic fields. The constructed $T$--$H$ diagram for $H\parallel y$ is shown in Fig.~\ref{fig:pd}(a). 
The shape and position of a boundary surrounding the ferromagnetic phase 
closely resembles the experimental magnetic phase diagram of URhGe  \cite{Gourgout16}. 
As soon as an applied field rotates in the easy plane away from the $b$ axis, the $c$ component of the field
couples linearly  to the order parameter and a  wing-shaped phase diagram emerges 
from the tricritical point  \cite{Griffiths73}. For URhGe such a phase diagram has been suggested
in a number  of experimental studies \cite{Levy07,Nakamura17}. In particular it has been argued that the quantum critical
point, which terminates the line of first-order magnetic transitions in a tilted field, plays a key role
for the field-induced superconducting phase \cite{Levy07}.

Combining the Monte Carlo simulations with the energy minimization results of Sec.~IIIB, we 
constructed the $T$--$H_b$--$H_c$ phase diagram of URhGe, see Fig.~\ref{fig:pd}(b). 
Essentially, we performed isothermal field scans for various values of the tilting angle $\alpha$ and 
extrapolated the magnetization jumps to zero in order to determine the critical angle $\alpha^*(T)$. 
Taking into account significant simulation times and multiple intermediate values of $\alpha$ to be explored,
we were able to complete this procedure only for two temperatures between $T=T_{\rm tr}$ and $T=0$.
The obtained phase diagram resembles qualitatively the expected behavior 
\cite{Griffiths73,Levy07,Nakamura17}. However, the limited number of points on the critical line
$\alpha^*(T)$ does not allow us to verify a tangential crossing of the transition lines, which is a general
property of the wing-shaped diagrams discussed in \cite{Taufour16}.

%%%%%%%%%%%%%%%%%%%%%%%%%%%%%%%%%%%%%%%%%%%%%%%%%%%
\section{Discussion}
%%%%%%%%%%%%%%%%%%%%%%%%%%%%%%%%%%%%%%%%%%%%%%%%%%%

We have formulated and investigated a minimal spin model  that  
accounts for all basic properties of the heavy-fermion ferromagnet URhGe. 
The experimental values of the Curie temperature
$T_C$, the reorientation field $H_m$, and the height of the magnetization jump have been used to fix
the spin Hamiltonian parameters. For the chosen set of parameters we were able to  
{\it quantitatively} reproduce the other experimental features such as the critical tilting angle  $\alpha^*$ for 
 $H\parallel bc$ plane, the position of the tricritical point
$(T_{\rm tr}, H_{\rm tr})$, and the whole $T$--$H_b$ phase diagram. This quantitative agreement 
supports our initial assumption that the magnetic subsystem in URhGe is described by 
a local moment model as opposed to the itinerant ferromagnetism scenarios. 

The theoretical description of the magnetic phase diagram  established in our work shows
that URhGe has a moderately weak in-plane anisotropy $E/J\sim 0.22$. As a consequence, 
a gapped magnon mode representing transverse spin oscillations must be present in 
the ferromagnetic state.  Further calculations of
the dynamical magnetic susceptibility $\chi^{\alpha\beta}({\bf q},\omega)$
for URhGe can be performed on the basis of the Hamiltonian
(\ref{H}) using the precessional equation of motion for local magnetic moments.
Information on $\chi^{\alpha\beta}({\bf q},\omega)$  is essential  
for the spin-fluctuation mechanism of the unconventional superconductivity, 
see \cite{Monthoux07} and 
references therein. % as well as \cite{Karchev03,Mineev11,Hattori13,Bulaevskii19} 
Previously, a microscopic theory of the triplet superconducting phases in URhGe 
has been developed using a toy model for the magnetic subsystem
\cite{Hattori13}. Our study opens a way for more realistic theoretical calculations for the reentrant superconducting
phase. Let us also mention that a promising route for deriving effective spin
Hamiltonians similar to (\ref{H}) has recently been discussed in the framework of the 
underscreened Kondo lattice model \cite{Scott24}.

Overall,  theoretical insights gained here can contribute to a broader understanding of 
magnetism in other uranium materials that exhibit metamagnetic transitions such as
UIrGe \cite{Yoshii06}, UCoAl \cite{Matsuda13}, and UTe$_2$ 
\cite{Miyake19}. In particular, UTe$_2$ has recently been the focus of 
numerous experimental and theoretical studies  \cite{Aoki22,Lewin23}.
In a spectacular parallel with 
the behavior of URhGe, UTe$_2$ remains superconducting up to a metamagnetic transition at 
$H_m=35$~T ($H\parallel b$) with a large magnetization jump of  $0.6\mu_B$ \cite{Miyake19}.
Magnetic properties of UTe$_2$ are dominated by relatively large local moments
$\sim 3\mu_B$ on U atoms  \cite{Lewin23}. However, the material
 remains paramagnetic down to the lowest studied temperatures. 
The inelastic neutron-scattering measurements unveil a complex pattern of coexisting antiferro- 
and ferromagnetic fluctuations in UTe$_2$ \cite{Knafo21}. Therefore, in addition to anisotropy, a realistic  spin model for 
this material should include competing frustrated exchange interactions. The development of the microscopic
description for UTe$_2$, as well as for other metamagnetic uranium materials, is undoubtedely an interesting open problem.

\acknowledgements
We are grateful to D. Aoki, D. Braithwaite, J.-P. Brison, A. Huxley, G. Knebel, F. L\'evy, V. P. Mineev, and T. Ziman
for useful and inspiring discussions and other help. The financial support was provided by the
 French Research Agency (ANR),  within the project FRESCO, Project No. ANR-20-CE30-0020.

% = = = = = = = = = = = = = = = = = = = = = = = = = = = = = = = = = = = = = = = = = = = = = = = = = = = 
\begin{figure}[t]
\centerline{\includegraphics[width=0.7\columnwidth]{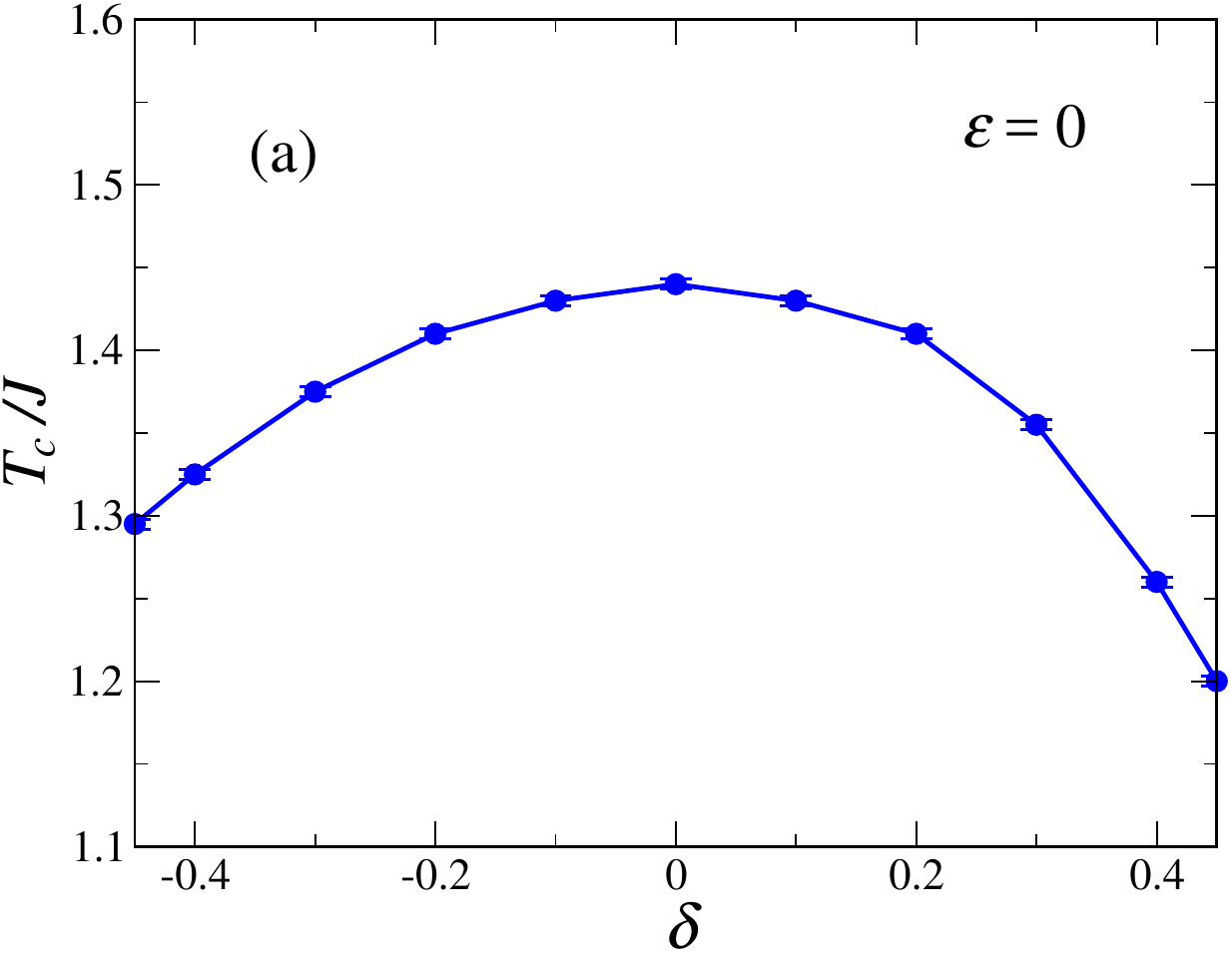}}    
\vskip 3mm
\centerline{\includegraphics[width=0.7\columnwidth]{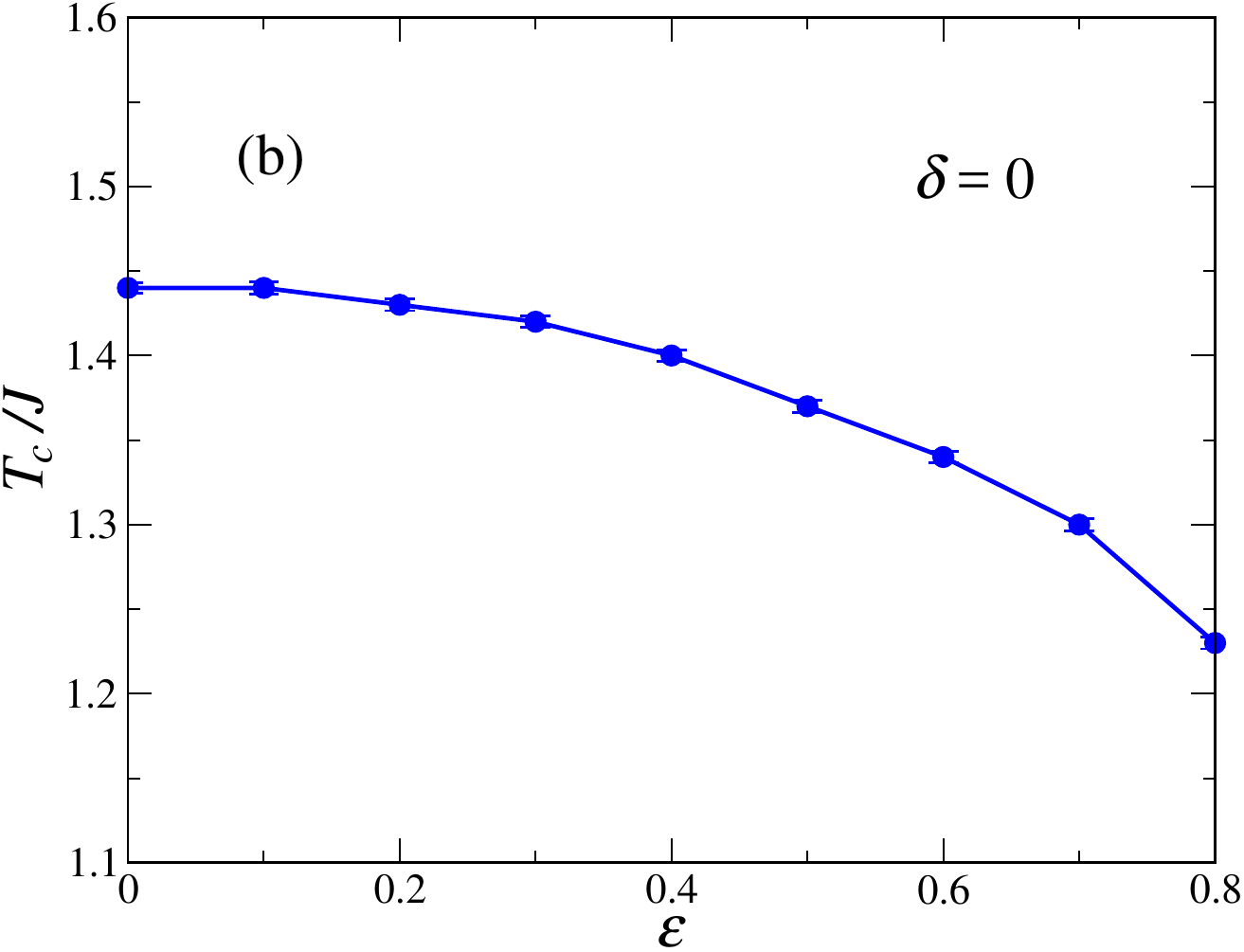}}
\caption{Dependence of the transition temperature of the classical Heisenberg spin model 
on an orthorhombic distortion (a)  $\delta$ and (b)  $\varepsilon$, see (\ref{A}) for notations.
}
\label{fig:de}
\end{figure}
% = = = = = = = = = = = = = = = = = = = = = = = = = = = = = = = = = = = = = = = = = = = = = = = = = = = 

%%%%%%%%%%%%%%%%%%%%%%%%%%%%%%%%%%%%%%%%%%%%%%%%%%%
\appendix
%%%%%%%%%%%%%%%%%%%%%%%%%%%%%%%%%%%%%%%%%%%%%%%%%%%

 % = = = = = = = = = = = = = = = = = = = = = = = = = = = = = = = = = = = = = = = = = = = = = = = = = = = 
\begin{figure}[b]
\centerline{\includegraphics[width=0.7\columnwidth]{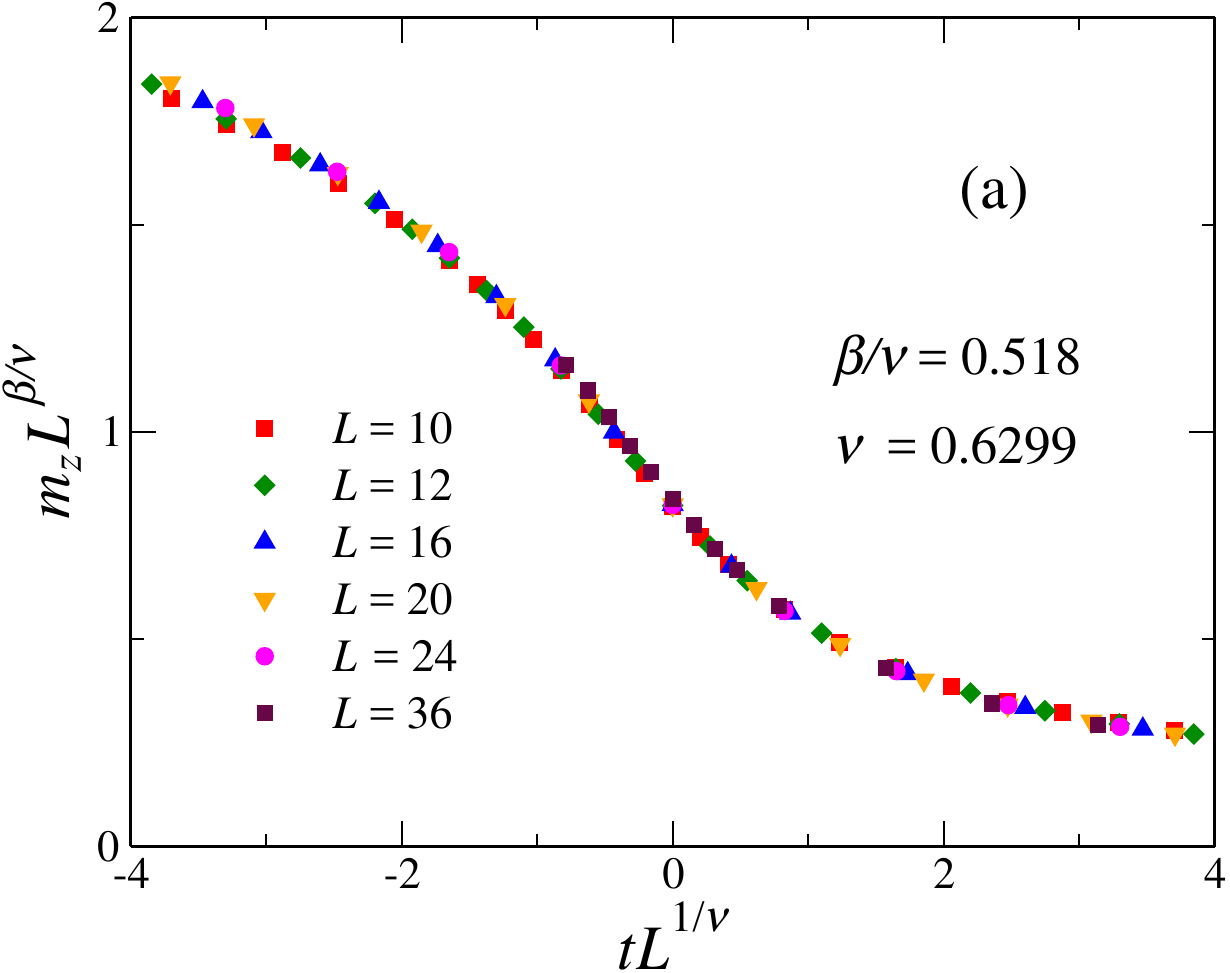}}    
\vskip 3mm
\centerline{\includegraphics[width=0.7\columnwidth]{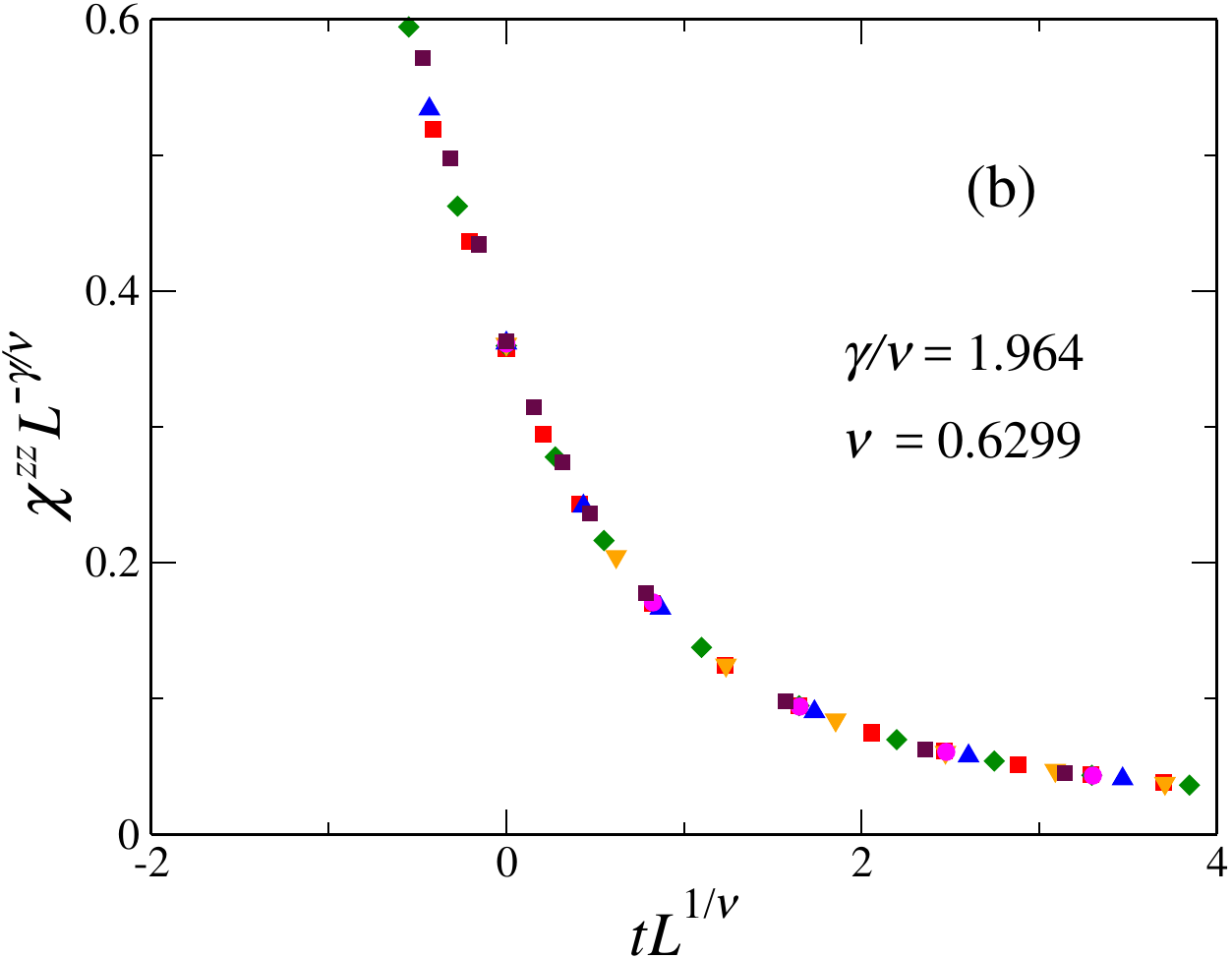}}
\caption{Zero field scaling plots  for the easy-axis magnetization 
$m_z$ (a) and the uniform susceptibility $\chi^{zz}$ (b)
 for different lattice sizes $L$ with the 3D Ising critical exponents, $t=(T-T_c)/T_c$ and
 $T_c/J = 1.88$.
%$\beta=0.3263$,  $\gamma=1.2371$, and  $\nu = 0.6299$.
Same symbols are used to represent lattices with the same linear sizes $L$ on both panels.
}
\label{fig:mchi}
\end{figure}
% = = = = = = = = = = = = = = = = = = = = = = = = = = = = = = = = = = = = = = = = = = = = = = = = = = = 

\section{}
%\section{Orthorhombic exchange}
In Appendix we include the  Monte Carlo data in support of the conclusions reached in the main text.

First, we consider an effect of orthorhombic pattern of exchange couplings   
on the transition temperature of a classical Heisenberg ferromagnet. For  a simple cubic lattice model 
with the nearest-neighbor 
exchange   $J$, the transition temperature is  $T_c = 1.4429J$ \cite{Chen93}. 
In an orthorhombic case, the exchange interactions may  differ along the three orthogonal directions:
\begin{equation}
\hat{\cal H}= - \frac{1}{2} \sum_{i,\boldsymbol{\rho}}J_{\boldsymbol{\rho}}\,\boldsymbol{S}_i \cdot 
\boldsymbol{S}_{i\pm \boldsymbol{\rho}} \ ,
\label{Hort}
\end{equation}
where $\boldsymbol{\rho} = x, y, z$. It is convenient to normalize the exchange constants to 
an average value $J=(J_x+J_y+J_z)/3$. The remaining `orthorhombic distortion' can be generally parametrized as
\begin{equation}
J_z/J = 1 + 2\delta \,, \quad
 J_{x,y}/J = 1 - \delta \pm \varepsilon \,.
 \label{A}
\end{equation}
Using the Monte Carlo simulations we have determined  the transition  temperature for two  slices of the surface  $T_c(\delta,\varepsilon)$: $\varepsilon=0$ and $\delta=0$.

Figure \ref{fig:de}(a) shows variations of the transition temperature between a layered ferromagnet with
$J_z :J_{x,y} = 0.1: 1.45$ ($\delta=-0.45$) and a model with weakly coupled chains $J_z:J_{x,y} = 1.9:0.55$
 ($\delta=0.45$).  In the second case, all three exchanges are different with the ratios
 $J_z :J_x:J_y = 1: 1.8:0.2$ for $\varepsilon=0.8$, see Fig.~\ref{fig:de}(b).  Overall, the highest transition temperature is always achieved for an
ideal cubic structure. Also, the quasi one-dimensional distortion of the exchange parameters
results in a stronger suppression of $T_c$ in comparison to the quasi two-dimensional pattern of exchanges.
However, the transition temperature variations do not exceed 10--20\%\ even for the limiting cases. Hence, the transition temperature of an orthorhombic ferromagnet, like URhGe, is mainly determined by an average exchange $J$ and
has a little dependence on a specific spatial distribution of $J_{\boldsymbol{\rho}}$.
 
Second, we consider the spin model of URhGe (\ref{H}) for the chosen set of the microscopic parameters
and present the zero-field scaling plots for the order parameter $m_z\propto  t^\beta$ and the magnetic susceptibility $\chi^{zz} = d m_z/dH_z\propto   t^{-\gamma}$ with $t = (T-T_c)/T_c$. 
The scaling theory predicts the following singularities for finite systems:
\begin{equation}
m_z =L^{-\beta/\nu} g_1 (t L^{1/\nu})\,, \quad \chi^{zz} =L^{\gamma/\nu} g_2 (t L^{1/\nu}) \,,
\end{equation}
where  $g_1(x)$ are $g_2(x)$ are the universal functions. The Monte Carlo simulation results are presented in Fig.~\ref{fig:mchi}
and show a nice collapse of the data from different clusters on the common curves for the 3D Ising set of
the critical exponents.

%%%%%%%%%%%%%%%%%%%%%%%%%%%%%%%%%%%%%%%%%%%%%%%%%%%

\end{document}